# Computational Explorations in Biomedicine: Unraveling Molecular Dynamics for Cancer, Drug Delivery, and Bimolecular Insights using LAMMPS Simulations

Reza Bozorgpour*

*Biomedical Engineering Department, University of Wisconsin-Milwaukee*
______________________________________

## Abstract

With the rapid advancement of computational techniques, Molecular Dynamics (MD) simulations have emerged as powerful tools in biomedical research, enabling in-depth investigations of biological systems at the atomic level. Among the diverse range of simulation software available, LAMMPS (Large-scale Atomic/Molecular Massively Parallel Simulator) has gained significant recognition for its versatility, scalability, and extensive range of functionalities. This literature review aims to provide a comprehensive overview of the utilization of LAMMPS in the field of biomedical applications.

This review begins by outlining the fundamental principles of MD simulations and highlighting the unique features of LAMMPS that make it suitable for biomedical research. Subsequently, a survey of the literature is conducted to identify key studies that have employed LAMMPS in various biomedical contexts, such as protein folding, drug design, biomaterials, and cellular processes.

The reviewed studies demonstrate the remarkable contributions of LAMMPS in understanding the behavior of biological macromolecules, investigating drug-protein interactions, elucidating the mechanical properties of biomaterials, and studying cellular processes at the molecular level. Additionally, this review explores the integration of LAMMPS with other computational tools and experimental techniques, showcasing its potential for synergistic investigations that bridge the gap between theory and experiment. Moreover, this review discusses the challenges and limitations associated with using LAMMPS in biomedical simulations, including the parameterization of force fields, system size limitations, and computational efficiency. Strategies employed by researchers to mitigate these challenges are presented, along with potential future directions for enhancing LAMMPS capabilities in the biomedical field.

**Corresponding Author***
Ph.D. Researcher, Biomedical Engineering Department, University of Wisconsin-Milwaukee, *E-mail:* bozorgp2@uwm.edu

# Introduction

Simulating the thermodynamic, mechanical, and chemical characteristics of solids and fluids rigorously, classical MD relies on an interatomic potential (force field). This potential defines the system's energy based on atom positions and other properties. Early applications, including studying radiation effects in solids and the dynamics of simple fluids, highlight the method's versatility [1-3]. MD has become widely used in physics, chemistry, biology, materials science, and related fields since its inception.

In nanotechnology filed such as water purification [4], MD also can play a crucial role in understanding the behaviour of nanoparticles at the atomic level, providing insights into their structural stability, surface properties, and interactions with surrounding molecules. It models systems as collections of particles, often atoms, and computes their time evolution by numerically integrating Newton's equations over numerous time steps. The forces on atoms are determined by the derivative of the analytic equations defining the potential function. This approach is computationally efficient, particularly for molecular liquids and solid-state metals, capturing electron mediated atomic interactions accurately. MD codes on standard workstations efficiently simulate systems with 10,000 to 1 million atoms, covering relevant length and time scales for significant physical and chemical phenomena over picoseconds to microseconds [5-8].

The surge in popularity of MD simulations can be attributed to their compatibility with the remarkable computational advancements driven by Moore's Law and extensive parallelism. Over the last few decades, both conventional CPUs and more recently GPUs have experienced substantial speed-ups. For instance, in 1988, an 8-processor Cray YMP achieved a Linpack speed of 2 gigaflops, while in 2012, a single IBM Blue Gene/Q CPU with 16 cores reached 175 gigaflops. The largest BG/Q machine, Sequoia, had nearly 100,000 CPUs. Anticipated GPU-based supercomputers in the next year or two are expected to reach exaflop ($10-18$) speeds, marking a half-billion-fold speed increase in just over 30 years for the most powerful supercomputers. This trend has also translated into speed enhancements for desktop machines and smaller clusters accessible to the wider scientific computing community [9, 10].

MD's computational efficiency stems from its per-time step cost scaling linearly as $O(N)$ for models with short-range interactions, a result of the finite number of neighbouring atoms within a specified cut off distance. Even for long-range Coulombic interactions, MD demonstrates effective scaling, with $O(N \log N)$ for FFT-based methods like particle-mesh Ewald or $O(N)$ for multipole methods (albeit with a larger pre-factor for the latter). Crucially, fundamental MD operations—including force computation, neighbour list construction, and time integration—are inherently parallelizable across atoms, ensuring optimal performance [11-13].

Advancements in computer hardware have significantly empowered MD simulations, unlocking the potential for expanded length or time scales. For systems with insufficient atoms to fully capitalize on larger machine parallelism, inventive algorithms have surfaced. These methods concurrently run multiple replicas of small systems, extending timescales, optimizing sampling efficiencies, or navigating the intricacies of free energy landscapes. Noteworthy strategies include accelerated MD methods, Markov state models, and enhanced sampling techniques, such as Meta dynamics.

In the realm of materials science applications, another factor driving MD's popularity is the accelerated evolution of hardware, inspiring the creation of more intricate and computationally demanding potentials. This, in turn, empowers more precise predictive modelling of material properties. Notable advancements include the formulation of many-body potentials like bond-order or reactive potentials, which come with computational costs several orders of magnitude larger than simpler pairwise additive potentials. More recently, the landscape has expanded to encompass machine learning (ML) potential. Unlike traditional physics-based equations, ML potentials rely on generic equations describing only the atomic neighbourhood's geometry. They are trained on extensive datasets derived from quantum calculations; a topic further discussed in later sections of this paper [14-16].

Driven by these considerations, a multitude of parallel MD codes has emerged, drawing in significant user communities. This encompasses codes specializing in bimolecular modelling, such as Amber, CHARMM GROMACS and NAMD, as well as those with a materials modeling focus, such as DL_POLY , HOOMD, and LAMMPS . Crucially, nearly all of these codes, LAMMPS included, are open-source and designed to support parallel computations on both CPUs and GPUs[17-20].

Bimolecular and materials MD codes differ in the types of models they support, despite some overlap. Biological systems display a richer variety of complex behaviors and phenomena compared to solid-state or soft non-biological materials. However, the interatomic potentials (force fields) employed for materials modeling are more numerous and diverse than those for bimolecular systems. This diversity arises from a wider range of chemical and physical interactions, thermodynamic conditions, and states of matter. Materials modeling encompasses various coarse-grained models, which operate not at atomic scales but at meso to continuum scales.

One key distinction lies in the size of systems and the time scales of interest, with considerable overlap. In the realm of biological MD, systems often revolve around one or a few solvated biomolecules, ranging from tens to hundreds of thousands of atoms. Intriguing phenomena, such as substantial conformational changes (e.g., protein folding), unfold at timescales that pose a challenge for MD modeling, spanning microseconds to milliseconds, or even seconds. On the flip side, materials MD frequently demands many millions or even billions of atoms to capture noteworthy phenomena. The rise in atom count proportionally reduces accessible timescales, potentially magnifying the relevance of the timescales from a physical standpoint. However, it's noteworthy that certain materials phenomena, like the shock response of solids, can be observed experimentally at relatively short picoseconds to nanoseconds timescales.

## A concise overview of LAMMPS' development

LAMMPS' inception dates back to the mid-1990s through a collaboration involving two US Department of Energy laboratories (Sandia National Laboratories and Lawrence Livermore National Laboratory) and three companies (Cray, DuPont, and Bristol-Myers Squibb). The objective was to develop a parallel MD code capable of harnessing spatial parallelism algorithms, as outlined in, to efficiently utilize the substantial processing power of large supercomputers, which, at the time, featured hundreds to a few thousand processors (cores in today's terms). The initial version, written in Fortran, was freely accessible, but users were required to navigate a perfunctory license agreement. Despite its availability, only around 100 users engaged over a decade, a phenomenon attributed to the deterrent effect of paperwork and legal complexities.

The rewriting of LAMMPS in C++ ensued because adding new models and features, like interatomic potentials, particle properties, constraints, and diagnostic computations, proved challenging in the Fortran based code. For performance optimization, the low-level Fortran-like data structures (vectors, arrays, simple structures) were preserved, and critical kernels were coded in a C-style manner. At a higher level, the object-oriented aspects of C++ were (and still are) employed to structure and organize the code, rendering it mostly "object-oriented C." It has been observed that this approach facilitates comprehension and modification of LAMMPS for users with diverse coding backgrounds. In 2004, the updated version was released as open-source code under the GNU General Public License (GPL), and within a month, it garnered more downloads than in the initial ten years

Over the following 18 years, the LAMMPS source code has expanded from 50 thousand to one million lines, incorporating contributions from several hundred users. The growing influx of contributions prompted the transition of the code repository and development workflow to GitHub in 2016. This move significantly bolstered our capacity to integrate and evaluate code contributions, rendering the code more robust. Leveraging GitHub facilities for bug reporting and automating configuration, regression, and unit testing further enhanced the development process.

Annually, LAMMPS witnesses a substantial user engagement with tens of thousands of tarball downloads, comparable GitHub accesses, and a vibrant community with several thousand interactions through mailing lists and forums—integral channels for user support. Notably, recent years have seen an expansion in licensing options, allowing companies and other interested parties to adopt LAMMPS under the GNU Lesser Public License (LGPL). This unique licensing flexibility facilitates the development of proprietary code utilizing LAMMPS as a library, encompassing applications such as GUIs, analyses, or workflow tools, all without the requirement to openly distribute the proprietary code. An illustrative instance is seen in DCS Computing, which integrates LAMMPS into its multiphysics simulation software tailored for particle simulations.

Since its inception, LAMMPS has been designed to harness distributed-memory parallelism through MPI. In terms of weak scaling, most of its models, with a minimum of a few hundred atoms per core, demonstrate scalability to millions of CPU cores. Over the past decade, LAMMPS has evolved to include GPU support, initially through CUDA and OpenCL code, and more recently, via the Kokkos library [21]. Kokkos, with back-ends compatible with GPUs from various vendors as well as Open MP, expands the spectrum of available options.

## Distributed computing strategies

This section outlines how various MD algorithms are structured in LAMMPS to facilitate parallelism across CPUs through MPI. Detailed discussions on parallelization in LAMMPS can be found in the following sections or the referenced papers. Throughout this and subsequent sections, the term "particles" is frequently used interchangeably with "atoms," although, as clarified in Section 1, particles may also represent finite size coarse-grained or continuum-scale objects.

*Partitioning:* The foundational spatial decomposition strategy employed by LAMMPS for distributed-memory parallelism through MPI was initially detailed in the original LAMMPS paper during the code's early development. While many concepts from that paper endure in the current code, they have undergone refinement in various aspects, as elaborated upon in this discussion.

The LAMMPS simulation box represents a 3D or 2D volume, which can take the form of an orthogonal or triclinic shape, as depicted in (Figure 1) for the 2D scenario. In an orthogonal configuration, the box edges align with the x, y, z Cartesian axes, resulting in all rectangular faces. Conversely, a triclinic box permits a more general parallelepiped shape, where edges align with three arbitrary vectors and faces become parallelograms. Each dimension can have periodic or non-periodic box faces, with fixed or shrink-wrapped boundaries. In the fixed case, atoms moving outside the face are deleted, while shrink-wrapped implies the box face continuously adjusts its position to enclose all atoms.

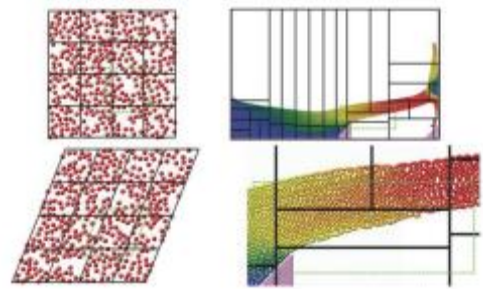

**Figure 1**. Different spatial decomposition of the simulation box for MPI parallelism.

The above picture is taken from. black lines show subdomain boundaries, and green dashed lines extend a processor's subdomain to include ghost atoms. Examples include orthogonal and triclinic boxes partitioned in the upper left and lower left, and variable-sized subdomains in the upper right, exemplified by an SPH model of water flow over a dam. The lower right provides a close-up view of one SPH processor subdomain. In the framework of distributed-memory MPI parallelism, non-overlapping subdomains fill the simulation box through spatial decomposition. Each of these subdomains, identified by solid lines in the figure, is allocated a distinct MPI rank or process.

In Fig. 1, the left-side images showcase the default partitioning with a regular grid of identical-sized subdomains. This works well for uniformly dense systems, but for models with varying density, load imbalance can hinder parallel efficiency. LAMMPS provides alternative partitioning options where subdomains can vary in size and shape, as long as they align with the simulation box faces—a framework utilized by all parallel algorithms. On the right side of (Figure 1), the images depict a partition generated by the Recursive Coordinate Bisectioning (RCB) algorithm. The RCB algorithm adapts subdomain size and shape, offering a solution for non-uniform density models and enhancing overall parallel efficiency.

*MPI communication:* In LAMMPS, processors store information for owned atoms and nearby ghost atoms, enabling the calculation of short-range interactions. This involves spatial decomposition of the simulation box, with each processor assigned a unique MPI rank. Ghost-atom communication is vital for calculations and occurs in two stages. For regular partitioning, atoms are exchanged first in the x-direction and then in the y-direction. In irregular partitioning, the communication pattern is more complex but follows a two-stage process. "Forward" communication sends attributes of owned atoms to neighbor processors, while "reverse" communication sends ghost atom attributes back to the owning processor. These operations are essential for various tasks within LAMMPS, such as exchanging coordinates and forces or sharing intermediate per atom values. Periodic box lengths are adjusted during forward communication across periodic boundaries, and the cutoff distance for exchanging ghost atoms is typically equal to the neighbor cutoff Rn. However, it can be longer if needed, exceeding the periodic box size or the diameter of a rigid body. If the cutoff distance extends beyond a neighbor processor's subdomain, multiple exchanges are performed in the same direction. In the irregular pattern, overlaps of a processor's extended ghost-atom subdomain with all other processors in each dimension are detected [22-23].

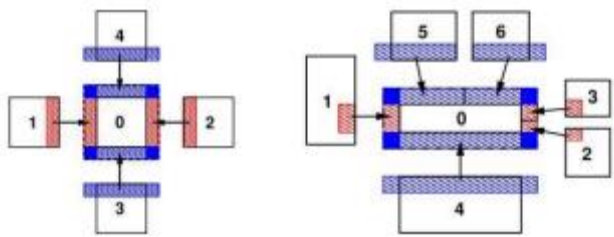

**Figure 2.** 2D simulation, regular vs. irregular partitioning, ghost atom communication illustration.

LAMMPS utilizes a Verlet-style neighbor list for efficient force computation. Each processor generates these lists, enumerating atom pairs within a cutoff distance. The neighbor list undergoes periodic rebuilding, and reneighboring is triggered by migrating atoms, with periodic boundary conditions enforced. For the linear O(M) time needed to build local neighbor lists, LAMMPS overlays a 3D (or 2D) grid of bins on the simulation domain. Each processor stores neighbor bins overlapping its subdomain, adapting to the neighbor cutoff distance. The "half" neighbor list optimally stores each I, J pair once, enhancing computational efficiency. LAMMPS supports various options for neighbor list builds. Users can adjust bin size, stencil attributes, and cutoff distances. The system accommodates hybrid models, handling multiple potentials and different cutoff lengths efficiently. Tailored algorithms address models with varying particle sizes and interactions, such as solvated colloidal systems or polydisperse granular particles [24, 25].

*Neighbor list:* In the pursuit of computational efficiency, processors in LAMMPS construct Verlet-style neighbor lists, meticulously enumerating atom pairs based on ownership and proximity. These lists, stored in a dynamic multiple-page structure, are pivotal for force computations. The periodic overhaul of these lists ensures accuracy. The computation relies on a cutoff distance, determined by a force cutoff ($R_f$) and a skin distance (s). Reneighboring is triggered when an atom moves half the skin distance. During reneighboring, atoms crossing processor domains undergo migration, and periodic boundary conditions ensure proper reassignment. The assembled atoms undergo spatial sorting for cache efficiency. Local neighbor lists are constructed linearly, overlaying the simulation domain with a grid of bins. The use of "half" neighbor lists, storing pairs only once, is a key strategy. This construction employs pre-computed stencils for efficient neighboring atom searches. Asymmetric stencils accommodate the unique storage needs of each atom in a half neighbor list.

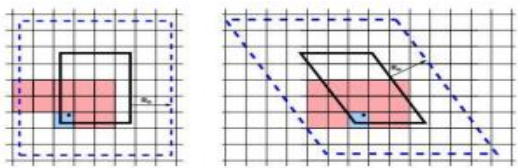

**Figure 3**. In 2D simulation, a processor's subdomain, outlined by thick lines, extends with a neighbor cutoff distance ($R_n$). The grid of neighbor bins, not bound by subdomain limits, shifts in the triclinic case, overlapping a tilted parallelogram. Blue and red-shaded bins, acting as a stencil, undergo adjustments for neighbor searches.

The intricacies of neighbor list construction in LAMMPS unfold through the deployment of stencils, elucidated in the figure for half lists with a bin size of $1/2 R_n$. In 2D, there exist 13 red+blue stencil bins (15 for triclinic), while 3D introduces 63 bins, 13 in the origin-containing plane and 25 in each of the two planes above in the z direction (75 for triclinic). Triclinic stencils, due to their non-periodic nature, incorporate additional bins. Adjustments in stencil and logic accommodate this peculiarity. The build process commences as each processor assigns its owned plus ghost atoms to neighbor bins, establishing a linked list of atom indices within each bin. A triply nested loop ensues, involving owned atoms I, the stencil bins around I's bin, and J atoms in each stencil bin. If $r_{IJ} < R_n$, J is appended to the list of I's neighbors. Further nuances emerge in LAMMPS's neighbor list options: Bin size, typically half of $R_n$, is a crucial choice. Smaller bins bring overhead; larger bins demand more distance calculations for atoms beyond the cutoff. Diverse potential commands may necessitate multiple lists with distinct attributes, including cutoff distances. Full lists, appearing twice for each I, J pair, and partial ghost-atom lists cater to specific potential needs. In hybrid models, a master list may partition into sub-lists for materials. Some models employ varied cutoff lengths for interactions between different particles in a single list. Solvated colloidal systems exemplify this, where interaction cutoffs differ dramatically. LAMMPS accommodates efficient algorithms for these scenarios, adapting stencils and bin sizes accordingly. The master neighbor list may partition into sub-lists, tailored for interactions within specific material subsets, streamlining computational efficiency. Notably, LAMMPS embraces algorithms accommodating diverse particle size ratios, ensuring robust neighbor list building for systems with substantial polydispersity [24, 25].

*Long range interactions:* Navigating the intricacies of long-range interactions within charged systems, LAMMPS employs the FFT-based Particle-Particle/Particle-Mesh (PPPM) method, as expounded in [26]. PPPM, a cousin of the Particle-Mesh Ewald (PME) method [27], offers nuanced distinctions, primarily in the charge interpolation to the mesh. For heightened accuracy, PPPM can operate on a slightly coarser grid than PME [28]. Additionally, LAMMPS extends support for alternative methods such as the Multilevel Summation Method (MSM) and the Fast Multipole Method (FMM), the latter facilitated through the ScaFaCos package [29, 30]. While MSM tends to be slower than PPPM, it accommodates fully non-periodic systems, unlike PPPM, which is constrained to fully periodic or non-periodic slab geometries. PPPM, akin to PME, overlays a regular FFT grid on the simulation domain, segregating Coulombic interactions into short- and long-range components. Real-space computation addresses the short-range segment through a loop over charge pairs within a cutoff distance, utilizing a neighbor list, as discussed in Section 3.3. Reciprocal-space computation, constituting the long-range component, involves several stages: charge interpolation to nearby FFT grid points, forward FFT, convolution operation in reciprocal space, inverse FFTs, and interpolation of electric field values to compute forces on each atom. The applicability of a single inverse FFT aligns with smoothed PPPM, while an analytical differentiation formulation requires three inverse FFTs. In the spatial decomposition partitioning depicted in (Figure 1), each processor assumes ownership of a brick-shaped portion of FFT grid points within its subdomain. Interpolation operations leverage a stencil of grid points surrounding each atom, necessitating the storage of ghost grid points around each processor's brick. Communication involves the exchange of electric field values on owned grid points with neighboring processors to update ghost point values, and vice versa for charge values. For triclinic simulation boxes, though FFT grid planes align with box faces, the mapping of values to/from grid points is executed in reduced coordinates, accommodating the tilted box conceptually as a unit cube. However, the larger FFT grid size required for a given accuracy in triclinic domains exceeds that for orthogonal boxes. Parallel 3D FFTs, acknowledged for substantial communication overhead, employ the pencil-decomposition algorithm [31], illustrated in (Figure 4). This algorithm interleaves computation and communication by transforming the layout of owned grid cells through successive stages, involving brick-to-pencil and pencil-to-pencil communication operations. This algorithm facilitates a scalable approach, with communication operations tailored to subgroups of processors. LAMMPS, by default, opts for point-to-point communication but provides the alternative of partitioned collective communication. This option optimizes FFT performance, notably demonstrated on the IBM Blue Gene/Q machine at Argonne National Labs, where partitioned collective communication outperformed other methods. The fftMPI library supports grid dimensions as multiples of small prime factors (2, 3, 5), accommodating any number of processors for FFTs. Though resulting decompositions may not align perfectly, subgroups for communication scale proportionally. For efficiency in performing 1d FFTs, the grid transpose operations illustrated in Fig. 4 also involve reordering the 3d data so that a different dimension is contiguous in memory. This reordering can be done during the packing or unpacking of buffers for MPI communication. At large scale, the dominant cost for parallel FFTs is often communication, not the computation of 1d FFTs, even though the latter scales as M log(M) in the number of grid points M. As discussed in Section 4.1, LAMMPS has an option to partition the allocated P processors into two subsets, say 1/4 and 3/4 of P in size. In this context, the smaller subset can perform the entire PPPM computation, including the FFTs, so that the FFTs run more efficiently (in a scalability sense) on a smaller number of processors. The larger subset is used for computing pairwise or many body forces, as well as intramolecular forces (bond, angle, dihedral, etc.). The two partitions perform their computations simultaneously; their contributions to the total energy, force, and virial is summed after they both complete. LAMMPS also implements PPPM-based solvers for other long-range interactions, dipole [32] and dispersion (Lennard-Jones) [33], which can be used in conjunction with long-range Coulombics for point charges. LAMMPS implements a GridComm class which overlays the simulation domain with a regular grid, partitions it across processors in a manner consistent with processor subdomains, and provides methods for forward and reverse communication of owned and ghost grid point values [34]. It is used for PPPM as an FFT grid (as outlined above) and for the MSM algorithm which uses a cascade of grid sizes from fine to coarse to compute long-range Coulombic forces. The GridComm class is also useful for models where continuum fields interact with particles. For example, the Two-Temperature Model (TTM) defines heat transfer between atoms (particles) and electrons (continuum gas) where spatial variations in the electron temperature are computed by finite differencing the discretized heat equation on a regular grid [35, 36]. The TTM was originally implemented in LAMMPS via a fix ttm command [37], and a new fix ttm/grid variant now uses the GridComm class internally to perform its grid operations in parallel.

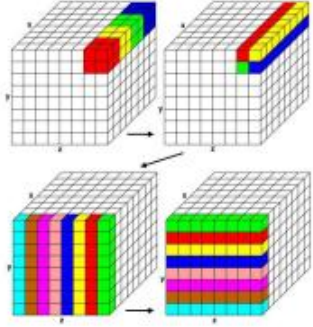

**Figure 4.** Parallel FFT on 64 processors for an $8 \times 8 \times 8$ 3D grid involves brick-to-pencil and pencil-to-pencil communication. Processors transform $2 \times 2 \times 2$ bricks into 1D pencils in the x dimension and transpose y and z dimensions in the grid, as shown in four diagrams.

## Aim of MD

Computer simulations are crucial for understanding how molecules interact and form structures. They complement experiments by revealing unique insights. MD stands out among simulation techniques, offering advantages over Monte Carlo (MC) by uncovering dynamic properties of systems. Simulations bridge the gap between microscopic interactions and macroscopic observations in the laboratory, providing accurate predictions of bulk properties within computational limits. They unveil hidden details behind measurements and aid in testing theories or models against experimental data. Simulations also allow exploration of extreme conditions, aiding in understanding phenomena that might be difficult or impossible to replicate in the lab. Ultimately, they aim to compare simulated results with experimental data, requiring robust models of molecular interactions like ab initio MD or facilitating investigations into generic phenomena without demanding perfect representations of the system.

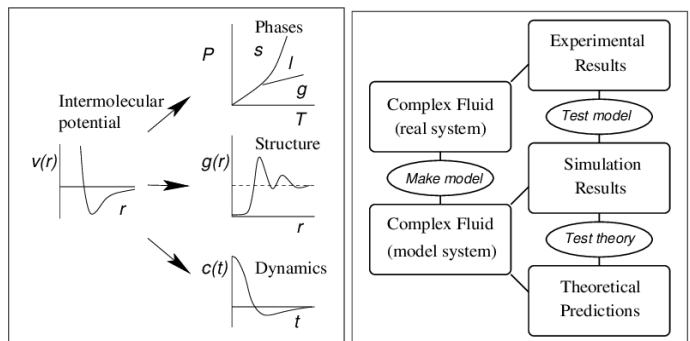

**Figure 5.** Simulations: Micro-Macro Bridge, Theory-Experiment Link.

## MD Interactions

A MD simulation involves solving classical equations of motion numerically, progressing through steps to describe the behavior of atomic systems. For a basic atomic setup, these equations govern the simulation process.

$$m_i \ddot{r}_i = f_i \quad f_i = -\frac{\partial}{\partial r_i} u \quad (1)$$

To achieve this objective, it is essential to compute the forces ($f_i$) that impact the atoms. These forces typically stem from a potential energy function, $u(r^N)$, wherein $r^N$ denotes the entirety of 3N atomic coordinates ($r_1, r_2, \dots, r_N$).

*Non-bounded-Interactions:* Traditionally, the segment of the potential energy $u_{non\_bounded}$, which accounts for interactions among atoms not bonded together, is conventionally divided into terms such as 1-body, 2-body, 3-body, and so forth.

$$u_{nonbounded}(r^N) = \sum_i u(r_i) + \sum_i \sum_{j>i} v(r_i, r_j) + \dots \quad (2)$$

In simulations of bulk systems with complete periodicity, the u(r) term, representing an external potential field or container wall effects, is typically omitted. Emphasis is placed on the pair potential $(r_i, r_j) = v(r_{ij})$, often neglecting higher-order interactions like three-body interactions. There exists extensive literature detailing how these potentials are determined experimentally or theoretically modeled [38-41]. In studying complex fluids through simulations, using simplified models that accurately reflect essential physics suffices. This chapter's emphasis lies on continuous and differentiable pair potentials, although discontinuous potentials like hard spheres and spheroids have also played a significant role. Among these, the Lennard-Jones potential stands out as the most commonly employed form[42].

$$v^{LJ}(r) = 4\varepsilon\left[\left(\frac{\sigma}{r}\right)^{12} \left(\frac{\sigma}{r}\right)^6\right] \quad (3)$$

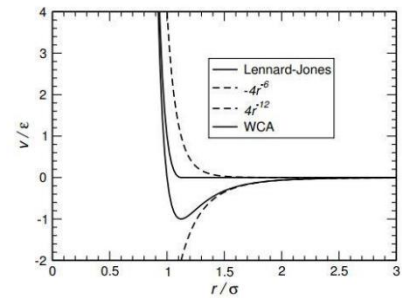

**Figure 6.** Lennard-Jones potential: $10-12, 10-6$ contributions and WCA shifted repulsion.

The potential relies on two parameters: σ (diameter) and ε (well depth) and was used in early studies of liquid argon properties. When excluded volume effects are more crucial than attractive interactions, truncating the potential at its minimum and shifting it upwards creates the WCA model [43]. If electrostatic charges are present, the relevant Coulomb potentials will be included.

$$v^{coulomb}(r) = \frac{Q_1 Q_2}{4\pi\varepsilon_0 r} \quad (4)$$

Effectively managing long-range forces in simulations, especially concerning polyelectrolytes, is a crucial aspect addressed in a later chapter by Holm [44]. This involves understanding the handling of charges ($Q_1$ and $Q_2$) and the permittivity of free space ($\varepsilon_0$), which are fundamental elements in these simulations.

*Bounding Potential:* Molecular systems are built using site-site potentials similar to Eq. (3). Electron density in a molecule is typically estimated through a single-molecule quantum-chemical calculation, represented by either partial charges using Eq. (4) or more precise electrostatic multipoles [45]. Intramolecular bonding interactions are crucial, thus incorporated into the simplest molecular models through related terms.

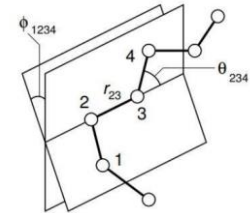

**Figure 7.** Illustration of the geometry of a basic chain molecule, defining interatomic distance $r_{23}$, bend angle $\theta_{234}$, and torsion angle $\varphi_{1234}$.

$$u_{intermolecular} = \frac{1}{2} \sum_{bonds} k_{ij}^r (r_{ij} - r_q) \quad (5a)$$

$$+ \frac{1}{2} \sum_{bend\ angles} k_{ij}^\theta (\theta_{ijk} - \theta_{eq}) \quad (5b)$$

$$+ \frac{1}{2} \sum_{torsion\ angles} \sum_m k_{ijkl}^{\emptyset,m} \left(1 + \cos(m\emptyset_{ijkl} - \gamma_m)\right) \quad (5c)$$

**Figure 7** depicts geometry. The "bonds" usually represent the separation $r_{ij} = |r_i - r_j|$ between adjacent atom pairs within a molecular structure. In Equation (5a), a harmonic form with a specified equilibrium separation is assumed, yet this isn't the sole option. The "bend angles" $\theta_{ijk}$ occur between successive bond vectors like $r_i - r_j$ and $r_j - r_k$, thus encompassing three atom coordinates.

$$\cos\theta_{ijk} = \hat{r}_{ij} \cdot \hat{r}_{jk} = (r_{ij} \cdot r_{ij})^{-1/2} (r_{jk} \cdot r_{jk})^{-1/2} (r_{ij} \cdot r_{jk})$$

The standard practice involves representing $\hat{r}$ as $r/r$. Bending terms usually take a quadratic form concerning the angular displacement from equilibrium, indicated by Eq. (5b), though periodic functions are sometimes used. Torsion angles, $\emptyset_{ijkl}$, are defined through three interconnected bonds, encompassing four atomic coordinates.

$$\cos\emptyset_{ijkl} = -\hat{n}_{ijk} \cdot \hat{n}_{jkl}, \quad n_{ijk} = r_{ij} \times r_{jk}, \quad n_{jkl} = r_{jk} \times r_{kl} \text{ with } \hat{n} = n/n$$

Simulation packages' force-field specifications outline Eq. (5) with parameters such as strength constants. Molecular mechanics force-fields, exemplified by MM3 and MM4, extend Eq. (5) by incorporating cross terms for improved predictions [MM3, MM4: Quantum mechanics aid in refining these force-fields, while comparing simulation results with properties facilitates their development. Distinct force field families (AMBER, CHARMM, OPLS) cater to larger molecules, simplifying Eq. (5) and determining parameters through quantum calculations and phase data AMBER: CHARMM: OPLS [46-55].

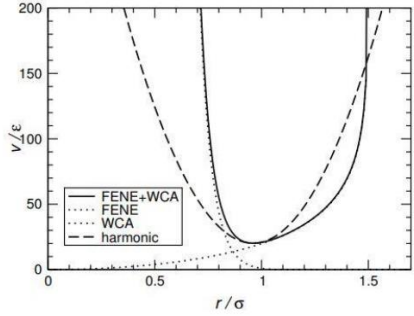

**Figure 8.** The FENE+WCA potential describes bonded atom interactions in a coarse-grained polymer chain, featuring both attractive (FENE) and repulsive (WCA) components.

*Force Calculation:* Once the potential energy function $u(r^N)$ is defined, the subsequent task involves computing the atomic forces.

$$f_i = -\frac{\partial}{\partial r_i} u(r^N)$$

Deriving atomic forces from site-site potentials is a relatively straightforward process. While handling the intramolecular part of the potential is a bit more complex, it mainly involves applying the chain rule. Appendix C of Reference [56] offers detailed examples elucidating this method. To illustrate, let's focus on a bending potential term for the polymer showcased in Figure 3, assuming it follows an expression like:

$$v = -k \cos\theta_{234} = -k(r_{23} \cdot r_{23})^{-1/2} (r_{34} \cdot r_{34})^{-1/2} (r_{23} \cdot r_{34})$$

These calculations will impact the forces acting on all three atoms. To compute these forces, the following components are required:

$$\frac{\partial}{\partial r_2}(r_{23} \cdot r_{34}) = r_{34} \quad \frac{\partial}{\partial r_3}(r_{23} \cdot r_{34}) = r_{23} - r_{34} \quad \frac{\partial}{\partial r_4}(r_{23} \cdot r_{34}) = -r_{23}$$

$$\frac{\partial}{\partial r_2}(r_{23} \cdot r_{23}) = 2r_{23} \quad \frac{\partial}{\partial r_3}(r_{23} \cdot r_{23}) = -2r_{23} \quad \frac{\partial}{\partial r_4}(r_{23} \cdot r_{23}) = 0$$

$$\frac{\partial}{\partial r_2}(r_{34} \cdot r_{34}) = 0 \quad \frac{\partial}{\partial r_3}(r_{34} \cdot r_{34}) = 2r_{34} \quad \frac{\partial}{\partial r_4}(r_{34} \cdot r_{34}) = -2r_{34}$$

Consequently,

$$\frac{\partial}{\partial r_2}\cos\theta_{234} = -r_{23}^{-1} r_{34}^{-1} (r_{34} - \frac{r_{23} \cdot r_{34}}{r_{23}^2} r_{23})$$

$$\frac{\partial}{\partial r_3}\cos\theta_{234} = -r_{23}^{-1} r_{34}^{-1} (\frac{r_{23} \cdot r_{34}}{r_{23}^2} r_{23} - \frac{r_{23} \cdot r_{34}}{r_{34}^2} r_{34} - r_{23} - r_{34})$$

$$\frac{\partial}{\partial r_4}\cos\theta_{234} = r_{23}^{-1} r_{34}^{-1} (\frac{r_{23} \cdot r_{34}}{r_{34}^2} r_{34} - r_{23})$$

Applying a similar approach to the torsional potential provides the forces acting on all four atoms involved in the torsional interaction.

# The MD Algorithm

Solving Newton's equations of motion does not immediately suggest activity at the cutting edge of research. The MD algorithm in most common use may even have been known to Newton [57]. Nonetheless, a rapid development in our understanding of numerical algorithms; a forthcoming review [58] and a book [59] summarize the present state of the field.

In a system of atoms with coordinates $r^N = (r_1, r_2, \ldots, r^N)$ and potential energy $u(r^N)$, atomic momenta $p^N = (p_1, p_2, \ldots, p^N)$ are introduced. The kinetic energy, $k(p^N)$, is expressed as $\sum_{i=1}^{N} |p_i|^2 / 2m_i$. The total energy, $H$, combining kinetic and potential terms, is given by $H = K + U$. The classical equations of motion can then be written as:

$$\dot{r}_i = p_i / m_i \text{ and } \dot{p}_i = f_i$$

The system involves a set of complex coupled ordinary differential equations characterized by both short and long timescales, necessitating adaptability in computational algorithms. Reducing the frequency of force calculations across atom pairs is crucial to manage computational expenses effectively. Advancing coordinates serves a dual purpose: accurately computing dynamic properties over usual timescales $\tau a$ and ensuring accuracy on the constant-energy hypersurface over longer times $\tau_{run}$ to capture a representative ensemble. Larger time steps are desired for quicker phase space sampling. Algorithms with lower order, avoiding high derivative storage, enable larger time steps for better energy conservation. Yet, accurately tracing the trajectory for long durations is impractical due to ergodic and mixing properties of classical trajectories. This supports variations of the Verlet algorithm, preferred despite historical predictor-corrector methods.

*The Verlet Algorithm:* Multiple versions of the Verlet algorithm exist, such as the original method and a 'leapfrog' form. However, our focus here centers on the 'velocity Verlet' algorithm [60].

$$p_i\left(t + \frac{1}{2}\delta t\right) = p_i(t) + \frac{1}{2}\delta t f_i(t) \quad (7a)$$

$$r_i(t + \delta t) = r_i(t) + \delta t p_i(t + \frac{1}{2}\delta t)/m_i \quad (7b)$$

$$p_i(t + \delta t) = p_i\left(t + \frac{1}{2}\delta t\right) + \frac{1}{2}\delta t f_i(t + \delta t) \quad (7c)$$

After executing step (7b), the algorithm evaluates $f_i(t + \delta t)$ during step (7c), enabling the advancement of coordinates and momenta using a time step $\delta t$.

The demo code for these steps is as follows,
```
for (int step = 1; step <= nstep; ++step) {
p = p + 0.5 * dt * f;
r = r + dt * p / m;
f = force(r);
p = p + 0.5 * dt * f;}
```

The Verlet algorithm possesses several crucial features: (a) exact time reversibility, (b) symplectic nature (to be elaborated on soon), (c) low temporal order enabling long time steps, (d) a single costly force evaluation per step requirement, and (e) straightforward programming. A fascinating theoretical derivation of this algorithm will be explained shortly.

*Limitations:* In classical computer simulations, intramolecular bonds are typically excluded from potential energy functions due to their high vibration frequencies. Instead, these bonds are constrained to maintain fixed lengths, managed through Lagrangian [61] or Hamiltonian [62] formalisms in classical mechanics. Algebraic relations, like a fixed bond length between specific atoms, aid in creating constraint equations and their time derivatives.

$$\chi(r_1, r_2) = (r_1 - r_2) \cdot (r_1 - r_2) - b^2 = 0 \quad (8a)$$

$$\dot{\chi}(r_1, r_2) = 2(v_1 - v_2) \cdot (r_1 - r_2) = 0 \quad (8b)$$

Within the Lagrangian formulation, the representation of constraint forces acting on the atoms will appear in this manner:

$$m_i \ddot{r}_i = f_i + \Lambda g_i$$

In this context, $\Lambda$ stands as the unspecified multiplier, while

$$g_1 = -\frac{\partial}{\partial r_1} = -2(r_1 - r_2) \quad g_2 = -\frac{\partial}{\partial r_2} = 2(r_1 - r_2)$$

The aim is to derive a solution for the multiplier Λ from equations expressing constraints. However, in practical simulations solved at discrete time steps, constraints are increasingly violated. A breakthrough occurred by proposing methods like SHAKE for the Verlet algorithm and RATTLE [63] for the velocity Verlet algorithm to accurately satisfy constraints at each time step. Formally, the aim is to solve a scheme by combining variables like $(r1, r2)$ as $r$, $(p1, p2)$ as $p$, for simplicity.

$$p\left(t + \frac{1}{2}\delta t\right) = p(t) + \frac{1}{2}\delta t f(t) + \lambda g(t)$$

$$r(t + \delta t) = r(t) + \delta t p(t + \frac{1}{2}\delta t)/m$$

Assuming $\lambda$ as

$$0 = \chi(r(t + \delta t)) \quad (9a)$$

$$p(t + \delta t) = p\left(t + \frac{1}{2}\delta t\right) + \frac{1}{2}\delta t f(t + \delta t) + \mu g t + (t + \delta t)$$

Taking $\mu$ as

$$0 = \dot{\chi} r(t + \delta t), p(t + \delta t) \quad (9b)$$

Implementing step (9a) involves defining variables without constraints.

$$\bar{p}\left(t + \frac{1}{2}\delta t\right) = p(t) + \frac{1}{2}\delta t f(t), \bar{r}(t + \delta t)$$
$$= r(t) + \delta t \bar{p}\left(t + \frac{1}{2}\delta t\right)/m$$

Afterwards, solve the nonlinear equation for λ.

$$\chi(t + \delta t) = \chi\left(\bar{r}(t + \delta t) + \frac{\lambda \delta t g(t)}{m}\right) = 0$$

Proceed by substituting the obtained solution back into the equation

$$p\left(t + \frac{1}{2}\delta t\right) = \bar{p}\left(t + \frac{1}{2}\delta t\right) + \lambda g(t), r(t + \delta t)$$
$$= \bar{r}(t + \delta t) + \frac{\lambda \delta t g(t)}{m}$$

Handling step (9b) involves defining,

$$\bar{p}(t + \delta t) = p\left(t + \frac{1}{2}\delta t\right) + \frac{1}{2}\delta t f(t + \delta t)$$

To find the second Lagrange multiplier μ, solve the equation accordingly.

$$\dot{\chi}(t + \delta t) = \dot{\chi}(r(t + \delta t)), \bar{p}(t + \delta t) + \mu g(t + \delta t) = 0$$

The equation, which is essentially linear due to the relationship $\dot{\chi}(\boldsymbol{r} + \boldsymbol{p}) = -g(\boldsymbol{r}) \cdot \boldsymbol{p}/m$, is solved first, followed by substituting the solution back.

$$p(t + \delta t) = \bar{p}(\delta t) + \mu g(t + \delta t)$$

The demo code is as follows,

```
for (int step = 1; step <= nstep; ++step)
{p = p + (dt/2) * f;
r = r + dt * p / m;
lambda = shake(r);
p = p + lambda;
r = r + dt * lambda / m;
f = force(r); p = p + (dt/2) * f;
mu_g = rattle (r, p);
p = p + mu_g;}
```

***Periodic Boundary conditions:*** In simulations with a small sample size, periodic boundary conditions are necessary to address surface effects. For instance, even in a system with 1000 atoms arranged in a $10 \times 10 \times 10$ cube, nearly half are on the outer faces, impacting measurements. Extending the simulation box with replicas resolves this issue. The minimum image convention enables atoms to interact with their closest counterparts. If an atom exits the base box, attention shifts to the incoming image for simulation. However, it's crucial to consider the artificial periodicity's impact, especially for properties affected by long-range correlations, notably in systems with extended potential ranges like charged or dipolar systems.

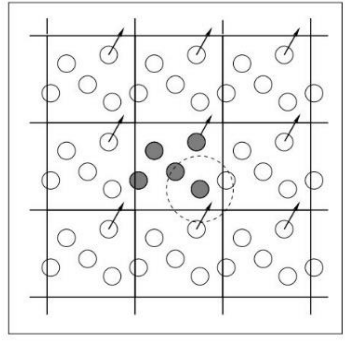

**Figure 9.** Schematic of Periodic boundary conditions.

***Neighbor List:*** In MD simulations, computing non-bonded interatomic forces involves numerous pairwise calculations, evaluating atom separations and interaction potentials. By setting a cutoff radius, $r_{cut}$, for the potential, the program skips force calculations for separations beyond cut, optimizing computation. However, this process remains time-consuming due to the significant number of distinct atom pairs (1/2 * N * (N - 1) for an N-atom system). Efficiencies are gained by using lists of nearby atom pairs. Verlet introduced a technique involving a cutoff sphere and a larger radius, $r_{list}$, forming a list of neighboring atoms within rlist. Only pairs within this list are checked during the force routine in subsequent MD steps. Periodic reconstruction of this list ensures accuracy, triggered by monitoring atom movements. For larger systems (N ≥ 1000), an alternate technique involves dividing the simulation box into regular cells, using linked lists to sort atoms into these cells for efficient neighbor searches within nearby and neighboring cells, optimizing the process for large systems with short-range forces.

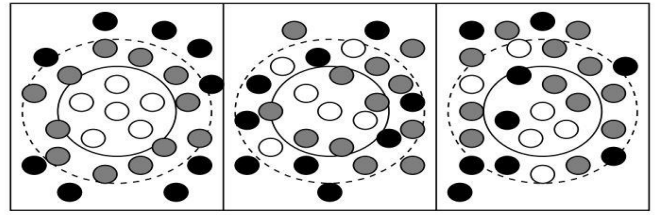

**Figure 10.** The Verlet list during construction, subsequent stages, and a delayed state.

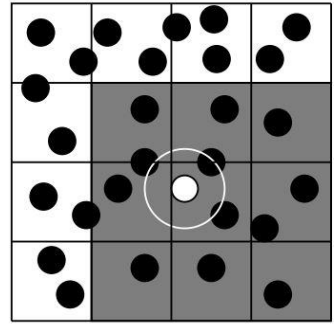

**Figure 11.** The cell structure with potential cutoff range shown. For neighbor search, only the atom's cell and its adjacent neighbors (shaded) are relevant.

***Time Dependence:*** Comprehending time-dependent statistical mechanics holds pivotal significance across three key domains in simulation. Primarily, some strides in grasping molecular dynamics algorithms stem from acknowledging the formal operator approach within classical mechanics. Secondly, a grasp of equilibrium time correlation functions, their association with dynamic traits, and particularly their correlation with transport coefficients, proves indispensable for aligning with experimental findings. Thirdly, the past decade has witnessed rapid strides in non-equilibrium molecular dynamics utilization, enhanced by a deeper comprehension of formal facets, particularly the interplay between dynamical algorithms, dissipation, chaos, and fractal geometry.

The evolution of the classical statistical mechanical distribution function, $\varrho(\boldsymbol{r}^N, \boldsymbol{p}^N, t)$, over time is governed by the Liouville equation. This equation is deduced straightforwardly by examining conventional Hamiltonian mechanics and analyzing the movement of representative systems within an ensemble across specific areas of phase space.

$$\frac{\partial \varrho}{\partial t} = -\left\{\sum_i \dot{\boldsymbol{r}}_i \cdot \frac{\partial}{\partial \boldsymbol{r}_i} + \dot{\boldsymbol{p}}_i \frac{\partial}{\partial \boldsymbol{p}_i}\right\} \varrho \equiv -iL\varrho \qquad (9)$$

The equation for ϱ involving the Liouville operator contrasts with the time evolution equation for dynamical variable $A(r^N, p^N)$, derived from applying the chain rule to Hamilton's equations. This comparison highlights differing approaches in understanding their evolution within classical mechanics.

$$\dot{A} = \sum_i \dot{r}_i \cdot \frac{\partial A}{\partial r_i} + \dot{p}_i \frac{\partial A}{\partial p_i} \equiv -iLA \quad (10)$$

The established solutions for the equations governing time evolution are

$$\varrho(t) = e^{-iLt}\varrho(0) \text{ and } A(t) = e^{iLt}A(0) \quad (11)$$

The exponential operator, termed the propagator, allows various manipulations. Analogies exist with fluid flow representations and quantum mechanics depictions, aiding in formulating classical response theory. This connects transport coefficients with both equilibrium and non-equilibrium simulations.

*Propagators and the Verlet Algorithm:* The derivation of the velocity Verlet algorithm involves a standard approximate breakdown of the Liouville operator. This breakdown preserves reversibility and is symplectic, ensuring conservation of volume in phase space.

Equation (11) features the Liouville operator, which can be

$$e^{iLt} = (e^{iL\delta t})^{n_{step}}_{approx} + \mathcal{O}(n_{step}\delta t^3)$$

$\delta t = t/n_{step}$ signifies small time segments. Inside the parentheses is an approximate propagator for short timesteps ($\delta t \to 0$), similar to how molecular dynamics divides long time spans into smaller steps using equation approximations. Splitting iL into two parts yields practical approximations,

$$iL = iL_p + iL_r \quad (12)$$

$$iL_p = \sum_i \dot{p}_i \cdot \frac{\partial}{\partial p_i} = \sum_i f_i \frac{\partial}{\partial p_i} \quad (13a)$$

$$iL_r = \sum_i \dot{r}_i \cdot \frac{\partial}{\partial r_i} = \sum_i m_i^{-1} p_i \frac{\partial}{\partial p_i} \quad (13b)$$

Taking the following approximation

$$e^{iL\delta t} = e^{(iL_p + iL_r)\delta t} \approx e^{\frac{iL_p \delta t}{2}} e^{iL_r \delta t} e^{\frac{iL_p \delta t}{2}} \quad (14)$$

As $\delta t$ tends towards zero, it approaches asymptotic exactness. For non-zero $\delta t$, it approximates $e^{iL\delta t}$ due to the non-commutativity of $iL_p$ and $iL_r$. Nevertheless, it retains perfect time reversibility and symplecticity.

We advance the equation of motion in steps, sequentially ignoring either the kinetic or potential part of the Hamiltonian. A derivation shows that each operator separately progresses the coordinates and momenta of the dynamic variable.

$$e^{iL_r \delta t} A(r, p) = A(r + m^{-1} p \delta t, p) \quad (15a)$$

$$e^{iL_r \delta t} A(r, p) = A(r, p + f \delta t) \quad (15b)$$

By using $r$ and $p$ instead of $r^N$ and $p^N$ to simplify, it becomes evident that the three steps in equation (14), utilizing the chosen operators, give rise to the velocity Verlet algorithm.

*Multiple Time steps:* An essential expansion of the MD technique enables its application in systems featuring multiple time scales. For instance, it can address molecules with intense internal springs (representing bonds) but softer external interactions. It also caters to molecules with markedly different short-range and long-range interactions or those composed of both heavy and light atoms. A basic MD algorithm must employ a sufficiently short time step to accommodate the fastest variables in these scenarios. Utilizing the Liouville operator formalism as described earlier, it's possible to create a time-reversible Verlet-like algorithm employing multiple time steps. In this scenario, the system comprises two kinds of forces: "slow" represented by $F_i$ and "fast" represented by $f_i$. The momentum obeys the equation $\dot{p}_i = f_i + F_i$. Subsequently, we decompose the Liouville operator as $iL = iIL_p + iL_p + iL_r$.

$$iIL_p = \sum_i F_i \frac{\partial}{\partial p_i} \quad 16(a)$$

$$iL_p = \sum_i f_i \frac{\partial}{\partial p_i} \quad 16(b)$$

$$iL_r = 3 \sum_i m_i^{-1} p_i \frac{\partial}{\partial p_i} \quad 16(c)$$

The propagator demonstrates an approximate factorization.

$$e^{iL\Delta t} \approx e^{iIL_p \Delta t/2} e^{i(L_p + L_r)\Delta t} e^{iIL_p \Delta t/2}$$

Here, $\Delta t$ denotes a long-time step. The intermediate section is further divided using the standard separation, iterating over small time steps $\delta t = \Delta t / n_{step}$:

$$e^{i(L_p + L_r)\Delta t} \approx [e^{iL_p \delta t/2} e^{iL_r \delta t} e^{iL_p \delta t/2}]^{n_{step}}$$

Fast-changing forces are computed frequently in short intervals, while slow-changing forces require only one calculation before and after this stage, done once per long time step.

*Rigid Molecule Rotation:* Including non-spherical rigid bodies in molecular models proves advantageous, particularly in simulating liquid crystals, colloidal systems, and polymers. This integration requires computation of intermolecular torques alongside forces and the implementation of classical dynamical equations to describe rotational motion. Such an approach allows for a more comprehensive representation of complex molecular systems, enabling simulations that capture the behavior of these diverse materials more accurately.

When intermolecular forces are represented as sums of site-site (or atom-atom) terms, transforming these into Centre-of-mass forces and torques around the Centre of mass can be straightforward. For instance, taking two molecules, A and B, with Centre-of-mass position vectors $R_A$ and $R_B$, respectively. By defining the intermolecular vector $R_{AB} = R_A - R_B$, and assuming the interaction potential can be expressed as such,

$$v_{AB} = \sum_{i \in A} \sum_{j \in B} v(r_{ij})$$

Considering 'i' and 'j' as atomic sites within their respective molecules, the computation then proceeds as:

The force acting on molecule A originating from molecule B:

$$F_{AB} = \sum_{i \in A} \sum_{j \in B} f_{ij} = -F_{BA}$$

The torque acting on molecule A originating from molecule B:

$$N_{AB} = \sum_{i \in A} \sum_{j \in B} r_{iA} \times f_{ij}$$

The torque acting on molecule B originating from molecule A:

$$N_{BA} = \sum_{i \in A} \sum_{j \in B} r_{jB} \times f_{ji}$$

Using $r_{iA} = r_i - R_A$ and $r_{jb}$ for site 'j' in molecule B, note that $N_{AB} \neq -N_{BA}$, but these equations offer direct calculations.

$$N_{AB} + N_{BA} + R_{AB} \times F_{AB} = 0 \quad (17)$$

Forces satisfying $f_{ij} = -f_{ji}$ along $r_{ij}$ ensure local angular momentum conservation due to the unchanged potential energy $v_{AB}$ during coordinate system rotations. However, in periodic boundaries, global angular momentum isn't conserved.

Another trend involves employing rigid-body potentials explicitly defined by center-of-mass positions and molecular orientations. The Gay-Berne [1] potential is one such example.

$$v_{AB}^{GB}(\hat{R}, \hat{a}, \hat{b}) = 4\varepsilon(\hat{R}, \hat{a}, \hat{b})[\varrho^{-12} - \varrho^{-6}] \quad (17a)$$

$$\varrho = \frac{R - \sigma(\hat{R}, \hat{a}, \hat{b}) + \sigma_0}{\sigma_0} \quad (17b)$$

This potential utilizes molecular axis vectors and center-to-center vector information, determining parameters for molecular size and orientation. Widely used in studying molecular liquids and crystals, it extends to non-uniaxial rigid bodies using orientation matrices for coordinate transitions. The following derivation, often infrequently presented, is detailed here [2]. In many instances, the pair potential can be expressed as:

$$v_{AB} = v_{AB}(R, \{\hat{a}_\alpha \cdot \hat{R}\}, \{\hat{b}_\beta \cdot \hat{R}\}, \{\hat{a}_\alpha \cdot \hat{b}_\beta\}) \quad (18)$$

In other words, the pair potential depends on the center-to-center separation R and all conceivable scalar products of the unit vectors $\hat{R}$, $\hat{a}_\alpha$, and $\hat{b}_\beta$. Employing the chain rule, we can express the force on A as:

$$F_{AB} = -\frac{\partial v_{AB}}{\partial R} = -\frac{\partial v_{AB}}{\partial R}\frac{\partial R}{\partial R} - \sum_{\hat{e}=\hat{a},\hat{b}} \frac{\partial v_{AB}}{\partial(\hat{e}.\hat{R})}\frac{\partial(\hat{e}.\hat{R})}{\partial R}$$

$$= -\frac{\partial v_{AB}}{\partial R}\hat{R} - \sum_{\hat{e}=\hat{a},\hat{b}} \frac{\partial v_{AB}}{\partial(\hat{e}.\hat{R})}\frac{\hat{e}-(\hat{e}.\hat{R})\hat{R}}{R} \quad (19)$$

The summation encompasses all orientation vectors on both molecules, denoted as $\hat{e} = \{\hat{a}_\alpha, \hat{b}_\beta\}$. Evaluating the derivatives of the potential is straightforward, assuming its general form as in Eq. (18). To compute torques, we used the approach in Let's examine the derivative of $v_{AB}$ concerning the rotation of molecule A by an angle ψ about any axis $\hat{n}$. By definition, this yields:

$$\hat{n}.N_{AB} = -\frac{\partial v_{AB}}{\partial \psi} = -\sum_{\alpha}\sum_{\hat{e}=\hat{R},\hat{b}} \frac{\partial v_{AB}}{\partial(\hat{e}.\hat{a}_\alpha)}\frac{\partial(\hat{e}.\hat{a}_\alpha)}{\partial \psi} \quad (20)$$

It aggregates combinations of unit vectors $(\hat{a}_\alpha, \hat{e})$, where $\hat{a}_\alpha$ rotates in tandem with the molecule, while the other, denoted by $\hat{e} = R$ or $\hat{b}_\beta$, remains fixed. This interplay results in a specific effect [3].

$$\frac{\partial \hat{a}_\alpha}{\partial \psi} = \hat{e}\times\hat{a}_\alpha \Rightarrow \frac{\partial(\hat{e}.\hat{a}_\alpha)}{\partial \psi} = \hat{e}.\hat{n}\times\hat{a}_\alpha = -\hat{n}.\hat{e}\times\hat{a}_\alpha$$

Equation (20) yields,

$$\hat{n}.N_{AB} = \hat{n}.\sum_{\alpha}\sum_{\hat{e}=\hat{R},\hat{b}} \frac{\partial v_{AB}}{\partial(\hat{e}.\hat{a}_\alpha)}\hat{e}.\hat{a}_\alpha \quad (21)$$

Selecting $\hat{n}$ as each of the coordinate directions successively enables the identification of every torque component:

$$N_{AB} = \sum_{\alpha}\sum_{\hat{e}=\hat{R},\hat{b}} \frac{\partial v_{AB}}{\partial(\hat{e}.\hat{a}_\alpha)}\hat{e}.\hat{a}_\alpha \quad (22)$$

The same logic for A and B

$$N_{AB} = \sum_{\alpha}\frac{\partial v_{AB}}{\partial(\hat{a}_\alpha.\hat{R})}\hat{R}\times\hat{a}_\alpha - \sum_{\alpha\beta}\frac{\partial v_{AB}}{\partial(\hat{a}_\alpha.\hat{b}_\beta)}\hat{a}_\alpha\times\hat{b}_\beta \quad (23a)$$

$$N_{BA} = \sum_{\beta}\frac{\partial v_{AB}}{\partial(\hat{b}_\beta.\hat{R})}\hat{R}\times\hat{b}_\beta + \sum_{\alpha\beta}\frac{\partial v_{AB}}{\partial(\hat{a}_\alpha.\hat{b}_\beta)}\hat{a}_\alpha\times\hat{b}_\beta \quad (23b)$$

It's notable that equations (19), (23a), and (23b) retain the relationship $N_{AB} + N_{BA} + R\times f_{AB} = 0$, as previously derived. When the potential cannot be easily expressed in terms of scalar products, a comparable derivation yields the expressions:

$$f_{AB} = -\frac{\partial v_{AB}}{\partial R} = -\frac{\partial v_{AB}}{\partial R}\hat{R} - \frac{\partial v_{AB}}{\partial \hat{R}}[\frac{1-\hat{R}\hat{R}}{R}] \quad (24a)$$

$$N_{AB} = -\sum_{\alpha}\hat{a}_\alpha\times\frac{\partial v_{AB}}{\partial \hat{a}_\alpha} \quad (24b)$$

$$N_{BA} = -\sum_{\alpha}\hat{b}_\beta\times\frac{\partial v_{AB}}{\partial \hat{b}_\beta} \quad (24c)$$

## LAMMPS in Biomedical Applications

***Previous work on molecular dynamics in cancer research:*** Our previous investigation was dedicated to the application of molecular dynamics simulations in the realm of cancer research. This investigation underscored the pivotal role of computational modeling techniques in unraveling the dynamic behaviors of biomolecules central to cancer pathways. Through these simulations, a comprehensive exploration of molecular interactions and dynamic changes within cancer-related bimolecular systems was achieved, contributing significant insights into the underlying molecular mechanisms governing cancer progression [66-67].

**Current Focus: Utilizing LAMMPS Molecular Dynamics in Cancer Research:**
In a study by Fu et al. [4], a Lennard-Jones type pair-potential method was implemented in LAMMPS to simulate coarse-grained lipid bilayer membranes. The method involved coarse-graining multiple lipids into a single particle in the thickness direction, allowing efficient simulations of large-scale lipid systems such as Giant Unilamellar Vesicles (GUVs) and Red Blood Cells (RBCs). The proposed method captured important mechanical properties of lipid bilayer membranes, including phase transitions, diffusion, and bending rigidity. Additionally, the effect of the cytoskeleton on lipid membrane dynamics and hydrodynamic interactions were considered by incorporating coarse-grained water molecules modeled with a Lennard-Jones potential. The simulations demonstrated the ability of the method to capture shape transitions and dynamics of vesicles and RBCs. The parallel computing capability of LAMMPS was utilized, showing promising results for large-scale realistic complex biological membrane simulations lasting over 1 ms.

Tan et al. [5] presented a parallel fluid-solid coupling model for simulating the interaction between viscous fluids and rigid or deformable solids. The model combines Palabos and LAMMPS, utilizing the Immersed Boundary Method (IBM) for coupling. The code is validated using various test cases, including ellipsoid particle motion, red blood cell stretching, and blood viscosity in tubes. It exhibits good scalability and performance in parallel computing. The developed model is applied to simulate the transport of flexible filaments (drug carriers) in flowing blood cell suspensions, showcasing its capabilities for biomedical applications.

The paper by Vachhani Savan et al. [6] focuses on investigating the potential of Bulk Metallic Glasses (BMGs) for cardiovascular applications, specifically as materials for cardiovascular stents. The study involves modeling a Cu-based BMG with compositional variations using multiscale methods and conducting simulations using LAMMPS software. The results reveal incremental changes in Young's modulus, ultimate tensile strength, and fracture mechanism through void formation in different BMG compositions. The study concludes that these BMGs demonstrate mechanical properties that meet the requirements for cardiovascular stents, suggesting their suitability for biomedical applications.

The study by Ye et al. [7] presents a multiscale and multiphysics computational method implemented in LAMMPS to investigate the transport of magnetic particles as drug carriers in blood flow. A hybrid coupling method handles the interfaces between Red Blood Cells (RBCs) and fluid, as well as magnetic particles and fluid. The method is validated using various behaviors and interactions in simple shear flow and under external magnetic fields. The proposed method is seamlessly integrated within LAMMPS, a parallelized molecular dynamics solver. Margination behaviors of magnetic particles and chains within blood flow are explored, showing that an external magnetic field can guide their motion and promote accumulation near the vascular wall. The computational method demonstrates high efficiency and robustness, providing a means to simulate nanoparticle-based drug carriers' transport in large-scale blood flow scenarios. The findings can contribute to the design of efficient drug delivery vehicles for improved imaging sensitivity, therapeutic efficacy, and reduced toxicity.

In the study by Oyewande et al.[8], the authors investigate the behavior of polyethylene (PE) through both experimental and computational methods. They focus on ion-beam sputtering of PE to understand its sputtering yield dependence, which is influenced by incident angles and ion energies. Molecular dynamics simulations are conducted on a PE system at a high temperature of 700 K to analyze its thermodynamic properties. The results reveal that the sputtering yield of PE depends on incident angles, with a peak observed at approximately 83°. The temperature of the PE system varies with time steps, while the structural and dynamical properties show minimal fluctuations as the polymer density increases. This study provides insights into the behavior of PE under ion-beam treatment and its thermodynamic properties at high temperatures.

In the study by Huilin et al. [9], they present a lattice model called magttice for hard-magnetic soft materials. This model partitions the elastic deformation energy and enables magnetic actuation through prescribed nodal forces. The researchers integrate magttice into the LAMMPS framework for efficient parallel simulations. They validate the model by examining the deformation of ferromagnetic beam structures and apply it to various smart structures such as origami plates and magnetic robots. The magttice model is also coupled with the Lattice Boltzmann Method (LBM) to study the swimming behavior of magnetic robots in water, resembling jellyfish locomotion. Overall, magttice offers an accessible and efficient platform for mechanical modeling and simulation, facilitating the design of magnetically driven smart structures.

J. Tranchida [10] provides an overview of LAMMPS, focusing on its features and capabilities, along with examples of modifications made to the software. The paper aims to offer a concise understanding of LAMMPS, which is a widely used software package for MD simulations. By exploring LAMMPS, the author provides insights into its various functionalities and highlights its versatility for customization. The paper includes specific examples of modifications made to the software, showcasing how LAMMPS can be tailored to suit specific research needs or to incorporate additional functionalities. Overall, the paper serves as a resource to gain a quick understanding of LAMMPS and provides practical examples of modifications that can be made to enhance its capabilities for MD simulations.

In the study by N. Ghamari et al. [75], the authors evaluate the interaction between curcumin and nigellin-1.1 with a brain antitumor molecule using molecular dynamic simulations. They employ DREIDING and universal force fields to model the atomic development of these compounds and calculate physical parameters such as total energy, center of mass distance, diffusion coefficient, and volume of the atomic structures. The results demonstrate an attractive force between curcumin and the brain antitumor structure, as well as between nigellin-1.1 and the brain antitumor structure. The study suggests potential tumor eradication effects based on changes in center of mass distances and increased volume of the brain antitumor structure following atomic interactions with curcumin and nigellin-1.1.

Covenery et al. in [76] provides an overview of MD simulation and its applications in various fields. The paper specifically focuses on the use of the LAMMPS software, which is a widely used MD simulator that allows for the analysis and visualization of the motion of atoms and molecules under various conditions. The authors discuss the basic principles of MD simulation and the methods used to determine the behavior of atoms and molecules at different temperatures and pressures. They also highlight the importance of reducing pre-steps in the simulation process to improve performance and portability. Overall, this paper provides a brief introduction to MD simulation and its application using LAMMPS, making it a useful resource for researchers and scientists interested in this field.

In a study conducted by Huilin Ye et al., a computational model was developed to explore the behavior of magnetic particles used in drug delivery within blood flow. Their method, integrated within the LAMMPS framework, allowed for robust and efficient large-scale simulations. The research focused on the influence of external magnetic fields on the transport and margination of these magnetic particles. Results demonstrated that the application of external magnetic fields could effectively guide the motion of these particles, enhancing their margination along the vascular wall. This work has significant implications for the design of more effective drug delivery systems, particularly in the context of targeted treatments for diseases like cancer. Additionally, the computational approach opens up possibilities for further investigations, such as drug carrier adhesion and diffusion within tumor tissues.

The purpose of the work by Hernan Chavez et al [77] is to address the challenge of transferring MD simulations between GROMACS and LAMMPS, two popular open-source software used for MD simulations. They introduce a Python 2 code called GRO2LAM, which provides a modular and open-source solution for translating input files and parameters from GROMACS to LAMMPS format. The study verifies the robustness of GRO2LAM by comparing simulation results obtained using GROMACS and LAMMPS after the format conversion. Three nanoscale configurations are examined, including a carbon nanotube, an iron oxide nanoparticle, and a protein in water, which are relevant to both engineering and biomedical fields. The results demonstrate good agreement between the energies obtained by the two different MD software, validating the effectiveness of GRO2LAM in achieving interoperability between these tools. This interoperability ensures the reproducibility of molecular dynamics models and allows for the easy exploration of their complementary capabilities and post-processing functionalities, potentially advancing the field of molecular dynamics simulations.

The work by Falick et al. [78] focuses on modeling confluent cell migration patterns in a particle dynamics framework using LAMMPS. They implement an active vertex model in LAMMPS, allowing forces to be applied to cell centers instead of vertices. The purpose is to simulate cell migration and understand its biophysical significance, which has applications in various fields, including organoid development, stem cell research, and endovascular device design. The study discusses the approach for implementing this model in LAMMPS and presents results for unjamming transitions, time step convergence, and cell migration replication for a monolayer of endothelial cells. However, some limitations are acknowledged, such as the computational cost of certain functions and the use of Forward Euler as a time integrator, which may affect simulation accuracy. The authors propose future work, including parallelization to handle larger and more complex simulations, implementing cell functionalities like growth, apoptosis, and division, solid state analysis to measure tissue elasticity, and extending the model into three dimensions. These advancements would enhance the model's capabilities and applicability in simulating cell behavior and interactions in various biological and engineering contexts.

In their study, Mohammad Pour Panah et al. [79] used MD simulations with the LAMMPS package to investigate the atomic interaction between Human Prostate Cancer's main protein (PHPC) and Fe/C720 Bucky Balls-Statin structures. The simulations were conducted under equilibrium conditions at 300 K and P=1 bar. The results showed that the interaction between Fe/C720 Bucky Balls-Statin and PHPC caused a decrease in the distance between them, ultimately leading to the destruction of PHPC. Notably, the volume of PHPC significantly increased during the simulations, which may have implications for the pharmaceutical industry. The study applied force fields like DREIDING and Universal Force Field (UFF) to estimate particle interactions. It's important to highlight that LAMMPS, a versatile MD package, was instrumental in conducting these simulations.

In this work by C.R. Grindon et al. [80], the authors focused on porting the AMBER forcefield to LAMMPS. Their aim was to enable simulations of DNA dynamics on extended timescales, which conventional algorithms could not achieve. To ensure the accuracy and efficiency of their porting, the researchers compared the results obtained using LAMMPS with their previous extensive analysis of DNA simulations conducted with AMBER. The study involved various aspects, including temperature control and performance scaling. The results demonstrated that LAMMPS effectively maintained the time-averaged behavior and dynamical characteristics of DNA observed in the AMBER simulations. The LAMMPS simulations showed great potential for probing previously unexplored regimes of dynamical behavior, opening new avenues for research in this field.

DNA methylation plays a critical role in regulating molecular dynamics within cells by influencing gene expression, chromatin structure, and protein-DNA interactions. Changes in methylation patterns, especially in regulatory regions like promoters, can alter the physical behavior and interactions of proteins, enzymes, and other molecules within the cell, thereby affecting molecular dynamics. For example, when methylation silences a gene, the lack of gene expression can change the abundance of proteins involved in key molecular pathways, altering cell behavior and dynamics [117]. In cancer, molecular dynamics are often disrupted by abnormal methylation patterns, which can lead to changes in the expression of oncogenes or tumor suppressor genes. Studies investigating the prediction of differentially methylated cytosines (DMCs) using computational methods can be applied to understand how these epigenetic modifications influence the overall molecular environment in cancer cells. By predicting and analyzing methylation changes, researchers can explore how these epigenetic factors may drive molecular dynamics, leading to altered cell behavior and potentially contributing to cancer progression. Thus, the connection between methylation studies and molecular dynamics lies in how these epigenetic modifications regulate and influence the interactions and behavior of molecules within cells, with particular relevance to cancer biology [118-119].

In the study by AlDosari et al. [81], the authors focused on drug release using nanoparticles to target cancer cells, particularly using functionalized graphene as a nanocarrier for the drug doxorubicin. MD simulations were employed to investigate intermolecular interactions and assess the suitability of graphene as a carrier. The results from the energy analysis, Gibbs free energy, hydrogen bond, radius of gyration, and Radial Distribution Function (RDF) indicated that graphene modified with an amine functional group is the most effective two-dimensional Nano carrier for transferring doxorubicin to cancer cells. This research highlights the potential of nanotechnology-based drug release systems in the selective recognition of cancer cells and overcoming limitations associated with conventional chemotherapy. The study's findings may pave the way for future laboratory investigations involving drug delivery using two-dimensional nanostructures, providing a promising avenue for cancer treatment.

Tsukanov et al. [82] focused on the interaction between two-dimensional aluminum oxyhydroxide nanosheets (AlOOH-NM) and cancerous cell plasma membranes. They aimed to understand the effects of these nanomaterials on cancer cell membranes. Using in silico simulations, the study estimated free energy changes and analyzed the orientation of Phosphatidylcholine (POPC) and Phosphatidylethanolamine (POPE) lipids when interacting with AlOOH Nano sheets. The results showed that the Nano sheets did not disrupt the cell membranes but were more likely to be adsorbed by the membrane surface, forming non-covalent bonds with the lipid head groups. This behavior could potentially lead to Nano sheet uptake by the cells through endosome formation. In summary, the research provides insights into the interaction between two-dimensional nanomaterials and cancer cell membranes, suggesting their potential for biomedical applications without disrupting the cell membrane's integrity.

The paper by [83] focuses on theranostics, a field aiming to improve disease diagnosis and targeted drug delivery, particularly in cancer treatment. It emphasizes the importance of specific molecular coupling between binding ligands and cancer cell receptors. Aptamers, genetic fragments with enhanced receptor specificity, are highlighted for their role in guiding drug molecules to specific sites, reducing harm to healthy cells. The paper explains the molecular interactions that drive aptamer-receptor binding and discusses the application of MD simulation to optimize this process for targeted cancer therapy. In conclusion, the paper underscores the global interest in developing tailored theranostic methods for various cancer types, as traditional treatments face limitations. Aptamers have emerged as promising tools for their receptor-driven binding efficacy. MD simulation is crucial for understanding and improving aptamer-receptor interactions in cancer therapy design.

In this study [84], the authors focus on cancer immunotherapy by targeting alternatively activated, M2-like Tumor-Associated Macrophages (TAMs), which play a significant role in tumor growth and metastasis. They design engineered nanoliposomes that mimic peroxidated phospholipids and are recognized by scavenger receptors on TAMs. Incorporating phospholipids with a terminal carboxylate group into the nanoliposome bilayers enhances their uptake by M2 macrophages with high specificity. MD simulations predict the "tail-flipping" of the sn-2 tail towards the aqueous phase, and docking data suggests interaction with Scavenger Receptor Class B type 1 (SR-B1). In vivo experiments demonstrate the specific targeting of these engineered nanoliposomes to M2-like macrophages, and when combined with specific inhibitors, they show promise in reducing the premetastatic niche and tumor growth. The study suggests that these "tail-flipping" Nano liposomes have potential as cancer immunotherapeutics for humans.

The paper, authored by Faizi et al., focuses on the application of the LAMMPS package in cancer research. Specifically, the study explores the potential of Silicon Carbide Nanotubes (SiCNTs) as carriers for targeted drug delivery of common anti-cancer drugs, including temozolomide, carmustine, and cisplatin, utilizing molecular dynamics simulations. The research investigates the effects of binding energy, drug distribution along the nanotube length, mean square displacement, body temperature, and zeta potential in the context of the drug delivery system's stability in the bloodstream. The findings reveal that cisplatin is not suitable for encapsulation in SiCNTs. While pure SiCNTs exhibit high stability in the bloodstream, their interaction energies with temozolomide and carmustine are insufficient to ensure drug retention. The introduction of carboxyl functional groups on nanotube surfaces neither significantly affects interaction energies nor enhances drug delivery system stability. In contrast, decorating the nanotube edges with hydroxyl groups results in strong interactions between temozolomide and SiCNTs, making them suitable for targeted drug delivery. The utilization of the LAMMPS package plays a crucial role in conducting these molecular dynamics simulations, providing detailed insights into the encapsulation process and the potential for cancer research applications. In conclusion, this study underscores the importance of LAMMPS in cancer research, specifically in assessing the potential of silicon carbide nanotubes, particularly those decorated with hydroxyl functional groups, as carriers for targeted drug delivery of certain anti-cancer drugs. This research offers a promising avenue for cancer treatment [85].

Sarah E. Johnstone et al. investigated the interplay between epigenetic and chromatin alterations in colorectal cancer and their impact on genome topology. By integrating topological maps of colon tumors and normal colons with epigenetic, transcriptional, and imaging data, the study revealed significant reorganization of genome compartments, particularly in tumors. This reorganization was accompanied by compartment-specific hypomethylation and chromatin changes. Notably, the research identified an intermediate compartment (compartment I) positioned between the canonical A and B compartments, which exhibited distinct epigenetic features and played a crucial role in tumor-associated changes. Importantly, these topological changes were not limited to cancer but were also observed in non-malignant cells with excessive divisions. Surprisingly, these changes were found to repress stemness and invasion programs while promoting anti-tumor immunity genes, potentially limiting malignant progression. The study challenges the conventional notion that tumor-associated epigenomic alterations are primarily oncogenic and sheds light on the potential tumor-suppressive effects of these topological alterations. The researchers employed molecular dynamics simulations using the LAMMPS software package to investigate genome organization. These simulations were conducted with reduced units, maintaining ensembles of genome organization at a constant temperature through Langevin dynamics. This comprehensive study has significant implications for understanding cancer progression and may offer insights into new strategies for early detection and therapeutic interventions in colorectal cancer and other related diseases [86].

In the study by Ostadhossein et al., the researchers tackle the challenge of poor solubility and variable drug absorption in cancer treatment, specifically STAT-3 inhibitors. They introduce a novel theranostics nanoplatform composed of luminescent carbon particles adorned with cucurbit uril to enhance the solubility of niclosamide, a STAT-3 inhibitor. This nanoplatform leverages host–guest chemistry between cucurbit uril and niclosamide, enabling the delivery of the hydrophobic drug, while carbon nanoparticles facilitate cellular uptake. The study confirms the successful synthesis of the Nano platform through physicochemical characterizations. Furthermore, the paper mentions that all the simulations in the study were conducted using the LAMMPS software package, implementing ReaxFF parameters. In vitro assessments in breast cancer cells demonstrate a significant twofold improvement in the drug's IC50, with efficient cellular internalization verified through various imaging techniques. In vivo experiments on mice with MCF-7 xenografts reveal a notable reduction in tumor size, coupled with immunohistochemistry evidence of STAT-3 phosphorylation downregulation. This research introduces a promising approach for STAT-3 inhibition in cancer therapy, combining host–guest chemistry, Nano carbon particles, and LAMMPS simulations to enhance drug solubility and efficacy [87].

In the study by Tohidi et al. MIL-100(Fe) was synthesized under biofriendly conditions, coated with Chitosan (CS), a natural polysaccharide, and investigated as a drug carrier for Cyclophosphamide (CP). The computational technique, using molecular dynamics software LAMMPS, was employed to predict drug loading in both MIL-100(Fe) and MIL-100(Fe)/CS. Powder X-ray diffraction analysis characterized the chitosan-coated MIL-100(Fe) loaded with cyclophosphamide (MIL-100(Fe)/CS/CP). Drug loading and release processes were quantified using UV spectroscopy at 193 nm. The toxic effect of MIL-100(Fe)/CS/CP was assessed on human breast cancer (MCF-7) cells. In vivo images and H&E analysis demonstrated the inhibition properties of MIL-100(Fe)/CS/CP on tumor cells. The research highlights the importance of computational calculations to gain insight into drug adsorption, providing a foundation for experimental investigations. The biocompatibility and anticancer properties of chitosan molecules were found to enhance the tumor inhibitory effect of the particles compared to MIL-100(Fe)/CP and free cyclophosphamide treatments [88].

In the study conducted by Xiaoyu Zhang et al., the authors present a label-free method for analyzing bodily fluids based on Surface-Enhanced Raman Scattering (SERS). The proposed technique involves large laser spot-swift mapping and electrochemically prepared silver nanoparticle substrates. This method enables the analysis of overall properties of multicomponent liquids, identifying low-concentration components, and was applied to serum-based cancer diagnosis. The large laser spot, formed by a scanning galvanometer, captures an average spectrum from different samples. Swift mapping then detects low-concentration components such as tumor biomarkers. The silver nanoparticle substrates exhibit strong SERS activity and wettability, enhancing adsorption and avoiding sample cluster formation. The method was tested for accuracy against mass spectrometry results and demonstrated sensitivity in detecting carcinoembryonic antigen in early colorectal cancer serum. The research emphasizes the potential application of this method for pan-cancer detection and staging, highlighting its advantages in sensitivity and reproducibility. Additionally, the study utilizes LAMMPS for statistical molecular bond length analysis and MATLAB 2018 for Principal Component Analysis (PCA). The results indicate the method's effectiveness in cancer diagnosis and suggest its broader application in medical liquid biopsies for rapid, accurate, and sensitive early cancer detection [89].

Researchers Jianchang Xu et asl. Zhang engineered copolymer-functionalized Cellulose Nanocrystals (CNCs) for a dual-function drug carrier triggering both pH-responsive and near-infrared (NIR)-induced drug release. Utilizing Poly(ε-caprolactone)-b-poly(2-(dimethylamino)-ethyl methacrylate) (PCL-b-PDMAEMA) through ring-opening polymerization and activators regenerated by electron transfer atom transfer radical polymerization (ARGETATRP), the CNC-based carrier encapsulates doxorubicin (DOX) and Indocyanine Green (ICG). This innovative drug carrier exhibits intrinsic pH response and NIR-triggered DOX release due to temperature-induced collapse of the PCL domain's crystal structure. Its rod-like morphology enhances cellular uptake, showing accelerated endocytosis compared to spherical counterparts. Simulations conducted in NVT ensembles using the LAMMPS package and DPD theory underline the CNCs' potential in cancer therapy, offering controlled release, biocompatibility, and enhanced efficacy through a combination of photo thermal and chemotherapy [90].

Li Xu, et al. developed a supramolecular cyclic dinucleotide nanodelivery system for STING-mediated cancer immunotherapy. Using endogenous molecules oleic acid and deoxycytidine, they created a ligand (3′,5′-diOA-dC) for the STING agonist c-di-GMP (CDG). This hydrophobic nucleotide lipid formed stable cyclic dinucleotide nanoparticles (CDG-NPs), enhancing CDG retention and intracellular delivery in the tumor site. With an average diameter of 59.0 ± 13.0 nm, CDG-NPs improved STING activation, TME immunogenicity, and anti-tumor immunity. Molecular dynamics simulations with LAMMPS supported the study's structural insights, offering a flexible CDN delivery platform for cancer immunotherapy [91].

Rossinelli et al. developed software for large-scale simulations of microfluidic devices, focusing on blood and cancer cell separation in complex channels with subcellular resolution. Using Dissipative Particle Dynamics (DPD) and GPU-accelerated supercomputers, they achieved unprecedented time-to-solution, outperforming current DPD solvers by 38X-45X. The simulations considered up to 1.43 Billion deformable RBCs, covering microfluidic device compartments. The software demonstrated potential applications in medical diagnosis and drug design, redefining the capabilities of multiscale particle-based simulations and contributing to the rational design of critical medical devices. Currently, software for DPD deployed on supercomputers is based on extensions of code originally developed for MD, such as LAMMPS [92].

Mercan et al. explore the critical buckling load of Boron Nitride Nanotube (BNNT), harnessing its exceptional electrical, physical, and mechanical attributes in comparison to Carbon nanotubes. Our approach involves two methodologies: applying Eringen's nonlocal elasticity theory for size-dependent critical buckling loads and utilizing LAMMPS software for molecular dynamics simulations to derive the critical buckling loads. The study centers on a Zigzag (5,5) BNNT comprising 400 atoms in MD simulation analyses. The findings showcase BNNT's outstanding mechanical, oxidation, hardness, corrosion resistance, and high-temperature durability. The critical buckling load is intricately linked to the length-to-diameter ratio (D/L) of the nanotube. Notably, MD simulations, with their capacity to model imperfect nanotubes, outperform size-effective continuum mechanics, particularly in stability evaluations [93].

In the research by Yarahmadi and Ebrahimi, the focus was on investigating the adsorption of paclitaxel (PTX), a widely used chemotherapy drug, onto graphene oxide (GO) surfaces in aqueous environments. The study aimed to understand the impact of hydrophilic functional groups, specifically epoxy (GO-O) and hydroxyl (GO-OH), on PTX adsorption. MD simulations, employing the LAMMPS software package, were utilized to evaluate adsorption energy and the average distance of PTX molecules from the graphene surface. The findings revealed a critical value (18%) for functional groups, beyond which the GO-OH system exhibited increased repulsion forces, leading to longer molecular distances due to compressive surface stresses. This study underscores the importance of functional group types in influencing PTX adsorption on GO surfaces within drug delivery systems, with implications for optimizing medical applications [94].

In the research by Bahreini et al., nanoscale vesicles known as nanosomes, formed from self-assembled nanosize components, were explored for drug delivery using the *Macrophage-Expressed Gene (MPEG-1) protein*. Employing the MD method, simulations were conducted in two phases. Initially equilibrated samples were assessed for drug delivery performance using parameters such as drug release ratio, root mean square displacement, charge density, and Zeta function. Computational outputs indicated the atomic stability of samples, and drug delivery was observed after 0.12 ns in an aqueous environment. Notably, the MPEG-1 based nanosome demonstrated a drug delivery ratio of 64%, suggesting its potential in clinical applications. The MD simulations were performed using the LAMMPS package, employing the Velocity-Verlet method for integration-based equations. The results highlighted the suitability of the DREIDING interatomic potential for determining the time evolution of the nanosome-drug system, emphasizing its stability and applicability in drug encapsulation/release processes for clinical treatment [95].

In the exploration by Roosta et al. through MD simulations, the intricate processes of encapsulating gemcitabine (GMC) within Boron Nitride Nanotubes (BNNTs) and subsequently releasing it were scrutinized. A noteworthy observation emerged, indicating GMC's inclination to position itself within the BNNT during encapsulation. For the release phase, the introduction of a new expelling agent, C48B12, demonstrated efficacy in liberating GMC from BNNT, primarily driven by robust van der Waals interactions. This novel avenue presents BNNTs as potential Nano-containers for drug delivery, showcasing the promising role of C48B12 in biomedical applications. The simulations were meticulously orchestrated using the LAMMPS software package under the Canonical (NVT) ensemble at T = 298K, employing a 1 fs time step over 10 ns [96].

In the investigation by DeLong et al., the immunological activity of physiological metal Oxide Nanoparticles (NP), particularly zinc oxide (ZnO) and cobalt oxide ($Co_3O_4$), was scrutinized in comparison with their complexes involving Anticancer Peptide (ACP) and RNA. The study delved into the impact of these compositions on cancer-associated or tolerogenic cytokines, revealing noteworthy distinctions. ZnO NP exhibited promising effects, influencing both molecular and cellular immunogenic activity. The examination extended to the incorporation of an anticancer RNA (ACR), Polyinosinic: Polycytidylic Acid (poly I:C), and an anticancer peptide (ACP), LL37, onto ZnO NP. Surprisingly, ZnO-LL37, but not ZnO-poly I:C or the tricomplex (ZnO-LL37-poly I:C), significantly suppressed IL-6 by over 98–99%. MD simulations validated the association of LL37 onto ZnO NP. The study advocates for further exploration of physiological metal oxide compositions, termed physiometacomposite (PMC) materials, in conjunction with ACP and/or ACR as a potential immuno-therapeutic against melanoma and other cancers. The simulations were executed using the LAMMPS software package under specific conditions, providing valuable insights into the immunological aspects of these nanostructures [97].

The study by Dehaghani et al. aimed to explore efficient Nano-based drug delivery systems for anti-cancer drugs, specifically investigating the encapsulation process of 5-FU within carbon nanotubes (CNT) and Boron Nitride Nanotubes (BNNT) using MD simulations. Utilizing LAMMPS software, the team employed the Tersoff potential to model interactions between atoms in the nanotubes. Through MD simulation at 300 K and 101.3 kPa pressure, they determined the storage capacity of BNNT (8,8) for 5-FU and quantified interaction parameters of the drug's atoms using the DREIDING force field. The simulations revealed that 5-FU was swiftly adsorbed into both CNT and BNNT, with a notably faster adsorption rate observed for the drug-BNNT complex due to stronger Van der Waals (vdW) interaction energies. Ultimately, the calculations of vdW interaction energy indicated decreased values at the end of the 15 ns simulation, reaching approximately −15 kcal·mol−1 for CNT and −45 kcal·mol−1 for BNNT, favoring the adsorption process. Free energy calculations supported spontaneous encapsulation within both nanotubes, with energies of −14 kcal·mol−1 for CNT and a more stable encapsulation of −25 kcal·mol−1 for BNNT. Additionally, the study determined that on average, six 5-FU molecules could be stably encapsulated within the BNNT cavity [98].

Yang et al. aimed to determine the most stable configuration of a single-stranded DNA aptamer (AptVEGF) targeting vascular endothelial growth factor (VEGF) using the stochastic tunneling-basin hopping (STUN-BH) method, followed by MD simulations to assess its thermal stability and interactions with VEGF. Employing the STUN-BH method in conjunction with LAMMPS, the research identified three primary configurations of AptVEGF/VEGF complexes. These configurations revealed the strong affinity of AptVEGF residues for VEGF surface loop fragments over alpha helix and beta sheet fragments. AptVEGF I predominantly occupied the VEGF loop fragment, while AptVEGF II adsorbed onto the remaining VEGF loop and beta sheet fragments, resulting in a 24.8% reduction in binding strength compared to AptVEGF I alone. Additionally, subsequent stable attachment of AptVEGF I and AptVEGF II allowed limited space for AptVEGF III to bind partially to VEGF, as confirmed by real binding experiments. The study further demonstrated the high sensitivity of the aptasensor constructed based on MD simulations, displaying a linear detection range of 10 pg/mL–10 ng/mL for VEGF. Through the STUN-BH method, the research not only identified the most stable AptVEGF configuration on VEGF but also elucidated the detailed interaction mechanisms between AptVEGF and VEGF. This approach uncovered the preferences of AptVEGF residues for specific VEGF fragments, shedding light on the design of highly sensitive aptamer-based biosensors for detecting VEGF concentrations [99].

Hatam et al. investigates the interfacial thermal conductance between gold and silica, crucial for enhancing the thermal properties of gold nanoparticles in cancer thermotherapy. Utilizing classical non-equilibrium MD, the study reveals that MD results differ from predictions by the conventional diffuse mismatch model. Specifically, the calculated interfacial thermal conductance between amorphous and crystalline silica and gold unveils intriguing differences. At 300 K, the conductance is approximately 61 MW/m2K for crystalline silica and around 134 MW/$m^2$K for amorphous silica. Additionally, variations in conductance concerning van der Waals interaction strength and temperature changes from 300 K to 700 K (a 30% increase) were observed. Notably, the study highlights the caution required when using the diffuse mismatch model, which predicted a conductance of about 140 MW/$m^2$K—more than double the molecular dynamics result. These findings hold promise for optimizing thermal applications involving silicate-coated gold nanoparticles, impacting cancer therapy and Nano fluid cooling systems. The interplay between molecular dynamics outcomes and conventional models underscores the complexity in accurately predicting interfacial thermal conductance (LAMMPS was employed for interfacial thermal conductance calculations) [11].

The paper authored by investigates the interaction of OH radicals, H$_2$O$_2$ molecules, and HO$_2$ radicals with DNA using reactive molecular dynamics simulations through the ReaxFF force field, facilitated by LAMMPS. The study delves into understanding the oxidative stress caused by OH radicals, showcasing their propensity to form 8-OH-adduct radicals—an initial step towards forming mutagenic markers for DNA oxidation. Notably, H$_2$O$_2$ molecules were found unreactive within the considered timescale. The simulations revealed the dynamic nature of OH radical interactions with DNA, including H-abstraction reactions, the formation of DNA oxidation products, and partial DNA strand openings in aqueous solutions. Key observations include OH radicals engaging in reactions leading to the formation of DNA oxidation markers, potential DNA strand breaks, and increased oxidation at exposed nucleotides over time. These findings shed light on OH radicals and indirectly H$_2$O$_2$ molecules as pro-apoptotic species, suggesting their role in controlled cell death and their potential use in cancer treatment via cold atmospheric pressure plasmas (CAPPs). The study highlights the significance of comprehending plasma species' roles in triggering tumor-specific apoptosis for selective anti-tumor treatment while preserving healthy cell integrity, emphasizing the need for further investigations in plasma-based oncology to elucidate their precise mechanisms in tumor-specific apoptosis [12].

The study by Shahbazi et al. focuses on elucidating the behavior of an Aluminum-based biosensor through all-atom molecular dynamics simulations using the ReaxFF potential via LAMMPS. By investigating surface properties and adsorption processes under varied flow conditions and target concentrations, the research aims to create a predictive model for biosensor surface properties. The findings demonstrate that changes in flow velocity significantly impact adsorption, with an increase in ethanol adsorption ranging from 7% to 80% as velocity varies from 0.001 m/s to 1 m/s. However, higher target molecule concentrations complicate this trend, prompting the need for future implementation of machine learning models for comprehensive biosensor performance prediction. Utilizing the ReaxFF potential validated through quantum mechanics data, this MD approach showcases its potential in studying concentration and velocity effects on the binding process of the biosensor. The results indicate that increased velocity correlates with decreased adsorption time and heightened adsorbed target molecules, particularly for cases with fewer than 100 Ethanol molecules. However, at higher concentrations, the challenge lies in identifying free-binding sites, prompting the proposal for future utilization of machine learning methods to optimize bio sensing technology design and predict sensor performance holistically [13].

Tsukanov et al. explored the interaction between Layered Double Hydroxide (LDH) nanosheets and cellular proteins, focusing on the potential impacts on cellular electrostatics and proposing novel avenues in anticancer medicine. Through unbiased MD simulations and the COPFEE approach, the research aims to understand the adhesion mechanisms of proteins onto LDH Nano sheets. The simulations, conducted under human body conditions, reveal insights into the interaction between glutamic acid, a common protein component, and the LDH Nano sheet surface in the presence of chloride anions. Utilizing the CHARMM27 force field and LAMMPS package, the study estimates the free energy of glutamic acid anion adsorption on Mg/Al-LDH Nano sheets. Remarkably, the simulations demonstrate that specific extracellular loops of voltage-dependent sodium channels form hydrogen bonds with the LDH surface in contact with cells. This observation suggests potential influences of LDH on the voltage-gated sodium channel, impacting its structure, dynamics, selectivity filter, gate groups, voltage-sensor domain motions, and channel entrance geometry. The findings present a rapid method for estimating small molecule adsorption free energy, potentially reducing computational costs. Moreover, this study identifies potential mechanisms through which LDH Nano sheets might affect the voltage-gated sodium channel, opening new directions for understanding their impact on cellular processes, particularly relevant in the context of anticancer therapy [103].

In the study conducted by Tsukanov et al., all-atom MD simulations were conducted using the LAMMPS and CHARMM27 force field to investigate the influence of a positively charged synthetic Nano film, replicating a fragment of the two-dimensional folded AlOOH structure, on a POPE/POPG lipid membrane. These simulations revealed that the charged Nano film exhibited an affinity toward the membrane, inducing membrane tightening and altering the orientation of lipid head groups. Notably, the Nano film's presence resulted in a perturbation of cation concentrations near the membrane surface, notably reducing Na$^+$ and K$^+$ ion concentrations. This interaction may potentially affect ion channel functionality within cellular membranes, suggesting plausible implications in biomedical applications, particularly in anticancer therapies. The study demonstrates the capability of the LAMMPS molecular dynamics package in elucidating intricate interactions between nanomaterials and biological membranes, shedding light on their potential biomedical significance [104-105].

The study by investigates the mechanistic underpinnings of therapeutic shock waves on cellular structures, crucial in understanding the impact of mechanical interventions on diseased and healthy cells. Employing atomistic simulations and a novel multiscale numerical approach, the research probes the biomechanical responses of representative membrane complexes under rapid mechanical loading akin to shock wave conditions. Contrary to expectations, the rupture characteristics of these complexes do not display substantial sensitivity to varying strain rates. Intriguingly, the presence of embedded rigid inclusions within cellular membranes appears to act as "mechanical catalysts," facilitating stretch-induced membrane disruptions while concurrently reinforcing the mechanical stiffness of the associated complexes. These findings offer valuable insights into potential biomechanical-mediated therapeutics, highlighting the role of rigid molecules in cellular membranes and their impact on mechanosensitive cellular processes. The subsequent simulations are conducted using LAMMPS71, utilizing initial configurations from well-equilibrated molecular systems. Lequieu et al. addressed the intricate relationship between histone modifications and the three-dimensional (3D) structure of eukaryotic genomes, focusing on the manifestation of this structure in gene expression. Existing computational tools often fall short in comprehensively illustrating how modifications on the scale of individual histones influence genome-wide structures spanning vast lengths. This work introduces a novel molecular model of chromatin, the 1CPN model, designed to bridge this gap. Using a rigorous multiscale approach, it integrates the detailed physics of nucleosomes, including histone modifications and DNA sequence, into a reduced coarse-grained topology. This model enables efficient kilo base-scale simulations of genomic DNA while accurately reproducing the free energies and dynamics of single nucleosomes and short chromatin fibers. Notably, the 1CPN model accommodates recent advancements in linker histone models. Moreover, this study utilizes the 1CPN model to investigate the influence of linker DNA on chromatin assembly, revealing strong dependencies on linker DNA length, pitch, and even DNA sequence. Implemented within the LAMMPS simulation package, the 1CPN model is made publicly available, offering a valuable tool for exploring the nuanced interplay between histone modifications and 3D genome organization [106-107].

Siani et al. investigated the interaction between dopamine-functionalized TiO2 nanoparticles and two overexpressed cancer cell proteins, PARP1 and HSP90, using atomistic MD simulations. These simulations focus on unraveling the interaction mechanisms by analyzing the protein residues in contact with the nanoparticles, the contact surface area, and variations in protein secondary structures under different pH and ionic strength conditions mimicking biological environments. Surface functionalization, NP charge, and solution conditions are considered, revealing that less acidic intracellular pH conditions, combined with cytosolic ionic strength, intensify PARP1 interactions with the nanoparticles while potentially weakening the HSP90 contribution. These findings provide a coherent explanation for experimental observations regarding the corona formation around nanoparticles in cancer cell cultures. All MD simulations were executed using the CHARMM implementation within the LAMMPS code. The comprehensive investigation of surface functionalization, NP charge, pH, and ionic strength conditions using the implicit-explicit solvent MD simulation framework shed light on the differential behavior of PARP1 and HSP90 within the protein corona. The observed variations in protein-nanoparticle interactions under different solution conditions significantly contribute to understanding the nuances of protein-nanoparticle interplay in biological environments and provide insights for future Nano medicine research aiming at therapeutic and diagnostic applications. Rocha et al. employed coarse-grained MD simulations to explore the interactions between gold nanoparticles and lipid bilayers, investigating the impact of hydrophobicity, charge density, and ligand length. The simulations reveal that hydrophobic and anionic nanoparticles exhibit limited interactions, while varying charge densities can lead to pore formation or wrapping of nanoparticles, resembling early stages of endocytosis. The interplay between charge density and ligand length emerges as a crucial factor in designing nanoparticles for drug and gene delivery applications. The utilization of LAMMPS software in building atomistic models has enabled these insights. Simulations propose that controlling surface chemistry can dictate different pathways for nanoparticle internalization into cells. Nanoparticles with low charge densities could facilitate an energy-independent translocation mechanism, forming nanoscale pores due to strong interactions between cationic terminals and phosphate groups in the lipid bilayer. Conversely, higher charge densities favor endocytic pathways, altering the structure of the lipid bilayer. Furthermore, the study suggests that shorter ligands and low charge densities may enhance penetration into cell membranes. These simulations offer detailed insights into nanoparticle-membrane interactions and outline design principles that influence distinct internalization mechanisms within cells [108-109]

Bao et al. explores the intricate delivery process of nanocarriers, addressing limitations in understanding this multistep process that hampers their efficacy in medicine. Six self-assembled nanocarrier variants with systematically altered physical attributes, such as size, shape, and rigidity, are detailed. Both in vitro and in vivo analyses assess their performance in blood circulation, tumor penetration, cancer cell uptake, and anticancer effectiveness. Additionally, the study introduces data and simulation-based models to comprehend how these physical properties, individually and collectively, influence each delivery step and the overall delivery process. The simulations utilized the LAMMPS code with periodic boundary conditions applied in three directions. The findings reveal that nanocarriers featuring a tubular shape, short length, and low rigidity exhibit superior performance compared to other variants. Moreover, the study provides theoretical models capable of predicting the individual and combined impacts of nanocarrier properties on the multistep delivery of anticancer therapies, enhancing the understanding of these complex processes [14].

Mohebali et al. investigates the mechanical properties of a hybrid carrier consisting of graphene sheets and carbon nanotubes loaded with doxorubicin (DOX) using MD simulations. The study shows that increasing the temperature reduces the carrier's elastic modulus. Chitosan inclusion enhances the carrier's mechanical characteristics. The research also explores the effects of graphene sheet size, aspect ratio, voids in the carbon nanotube structure, and the presence of DOX on the carrier's mechanical behavior. Additionally, the study examines how the physical traits of carbon nanotubes, such as size and chirality, impact the carrier's mechanical features. [111]

Zhang et al. investigated the thermal properties of EG–water–Au Nano fluids aimed at enhancing heat dissipation in high heat load internal combustion engines. MD simulations examined the impact of EG mass concentration on thermal conductivity, elucidating the microscopic mechanisms underlying thermal enhancement. Results demonstrate that higher water content positively correlates with thermal conductivity in EG–water solutions. The addition of Au nanoparticles boosts the thermal conductivity of these solutions by up to 17.20%. The simulations reveal that the introduction of nanoparticles doesn't significantly alter the conformation of the EG molecular chain but enhances van der Waals interaction energies in Nano fluid systems. The radial distribution function analysis highlights stratified adsorption layers of base liquid molecules on nanoparticle surfaces, mimicking solid-like microstructures and elevating Nano fluids' thermal conductivity. LAMMPS was employed for simulations in this study. The conclusions drawn include the positive correlation between water content and thermal conductivity, the minimal impact of temperature variation and nanoparticle addition on EG molecular conformation, and the nanoparticle-induced increase in van der Waals interaction energies. Further analysis of radial distribution function and density distribution indicates stratified adsorption layers of base liquid molecules on nanoparticle surfaces, resembling solid microstructures, contributing to heightened thermal conductivity in Nano fluids. These findings hold significance in advancing traditional alcohol–water coolants' heat transfer capabilities. Abadi et al. explored the fabrication of Nano pores in a single-layer graphene Nano sheet via MD simulation utilizing cluster bombardment. Investigating four distinct locations with varying properties, the study examines the impact of kinetic energy, cluster type, diameter, and incident points on Nano pore characteristics. Results demonstrate that controlling cluster type, diameter, and energy influences Nano pore size and quality. Si cluster bombardment yields larger Nano pores compared to SiC and diamond clusters, with the 2 nm diameter diamond cluster fabricating the most suitable Nano pores, exhibiting smoother edges and a consistent area size distribution. Additionally, the study investigates the effect of straining the Nano sheet on Nano pore topography, showing that tensile strains lead to larger Nano pores with smoother edges, while compressive strains produce more irregular topography. All simulations were performed using the open-source software LAMMPS. This research provides insights into Nano pore fabrication in graphene Nano sheets, highlighting the role of cluster type, diameter, incident points, and strain effects on Nano pore characteristics, crucial for advancing Nano pore-sequencing DNA technology [112-113].

The study by Siani et al. investigates the impact of cRGD ligand density on the structural-functional parameters of PEGylated TiO2 nanoparticles (NPs) for effective binding to αVβ3 integrin's, utilizing atomistic MD simulations. By varying cRGD densities conjugated to PEG chains on highly curved TiO2 NPs, the research reveals crucial insights into ligand presentation, stability, and conformation in explicit aqueous environments. Low densities favor an optimal spatial arrangement of cRGD ligands outside the PEGylated layer, while high densities lead to ligand over-clustering, driven by enhanced ligand-ligand interactions and reduced water accessibility. The findings emphasize that cRGD density modulation significantly influences the effective availability and arrangement of targeting ligands on NP surfaces, aligning with prior experimental investigations. Simulations demonstrate the spatial presentation of cRGD ligands at moderate densities (0.2-0.5 ligand nm$^{-2}$) being favorable for effective αVβ3 integrin binding, while high densities (> 1.0 ligand nm-2) hinder ligand solvation and presentation, impacting targeting efficiency and cellular uptake. This research offers crucial atomistic insights for designing more effective cRGD-based targeting nanosystems. The study, conducted using CHARMM in LAMMPS, provides in-depth insights into ligand density effects on NP surfaces, corroborating experimental observations and guiding the design of future cRGD-targeting Nano devices. Further explorations should delve into ligand-receptor binding and expanding the model's complexity to mimic real experiments, albeit potentially utilizing computationally less expensive approaches for larger systems [114].

The study delves into computational methodologies' increasing relevance in comprehending cell adhesion dynamics and their role in morphogenesis. The focus lies in examining how cells adapt to a granular bed concerning their size. The model, adapted from prior research by Cunha et al., introduces a fresh approach by employing LAMMPS, an open-source computational tool. Utilizing Langevin dynamics equations and a velocity Verlet scheme implemented within LAMMPS, the study explores the movement of particles and cell elements within the granular bed. The primary goal is to offer insights into cell behavior on granular surfaces, leveraging the precision and simulation capabilities of LAMMPS. The findings elucidate the cell's response to varying granular bed conditions based on relative size. However, despite the model's adaptation from previous work, the cell fate observed does not entirely align with the outcomes reported by Cunha et al. Yet, the study does highlight a transition observed in cell behavior between beds comprising significantly smaller cells than the granular particles. This transition demonstrates that smaller cells can enhance their adhesion and spread across the particles. Future improvements aim to study the effect of mixed mobile and non-mobile particle compositions on cell survival and refine the model to better match experimental observations [115-116].

## Conclusion

In summary, LAMMPS emerges as an indispensable and robust software package in the realm of MD simulations, specifically in biomedical research. Its unparalleled capabilities and versatile features render it an essential instrument for exploring intricate biological systems at the atomic scale. LAMMPS' scalability is a cornerstone, enabling researchers to efficiently simulate vast and complex systems, while its extensive repertoire of force fields and potential models accurately represents the diverse spectrum of bimolecular systems. The suite of comprehensive analysis tools embedded within LAMMPS is instrumental in deriving profound insights from simulation data, significantly augmenting our comprehension of multifaceted biological processes.

Furthermore, LAMMPS' adaptable and modular architecture empowers researchers to tailor the code, catering precisely to their unique research requirements. This adaptability facilitates the integration of supplementary functionalities and the innovation of new techniques, addressing the distinctive challenges encountered in biomedical investigations. The adaptability, resilience, and widespread utilization of LAMMPS have firmly established its status as a standard tool, enabling scientists across the globe to propel our comprehension of crucial biomedical phenomena. As the field continues to advance, LAMMPS is poised to retain its preeminent position in MD simulations, contributing significantly to groundbreaking discoveries and advancements in biomedical research.

## Motivation

This research is propelled by the burgeoning role of computational simulations, particularly in the realm of biomedical sciences. Harnessing the robust capabilities of LAMMPS for conducting molecular dynamics simulations, this study endeavors to harmonize multiple facets encompassing cancer research, drug delivery mechanisms, and bimolecular investigations. The primary aim is to employ computational tools to untangle the intricate behaviors inherent in biological systems, unravel the underlying mechanisms governing disease progression, optimize the intricacies of drug delivery methodologies, and decode the fundamental interactions within biomolecules. Through these endeavors, this research endeavors to bridge critical knowledge gaps in contemporary biomedical understanding, thereby illuminating potential pathways for the development of innovative therapies and strategic interventions to address multifaceted medical challenges.

## Conflict of interests
The corresponding author states that there is no conflict of interest.

## Funding
The author received no financial support for the review, authorship, and publication of this article.

## Ethics approval
This article is based on previously conducted studies and does not contain any new studies with human participants or animals performed by any of the authors.

## References

1. Gibson, J. B, et al. "Dynamics of radiation damage." *Phys Rev.* 120.4 (1960): 1229. [Google Scholar] [Crossref]
2. Rahman, Aneesa., "Correlations in the motion of atoms in liquid argon." *Phys Rev.* 136.2A (1964): A405. [Google Scholar] [Crossref]
3. Verlet, Loup. "Computer experiments on classical fluids. I. Thermodynamical properties of Lennard-Jones molecules." *Phys Rev* .159.1 (1967): 98. [Google Scholar] [Crossref]
4. Rafie, S.F., et al. "Hydrothermal synthesis of Fe3O4 nanoparticles at different pHs and its effect on discoloration of methylene blue: evaluation of alternatives by TOPSIS method." *Mater Today Commun.* 37 (2023): 107589. [Google Scholar] [Crossref]
5. Thompson, Aidan P., et al. "LAMMPS-a flexible simulation tool for particle-based materials modeling at the atomic, meso, and continuum scales." *Comput Phys Commun.* 271 (2022): 108171. [Google Scholar] [Crossref]
6. Perez, Danny, et al. "Accelerated molecular dynamics methods: introduction and recent developments." *Annu Rep Comput chem.* 5 (2009): 79-98. [Google Scholar] [Crossref]
7. Husic, Brooke E. & Vijay S. Pande. "Markov state models: From an art to a science." *J Am Chem Soc.* 140.7 (2018): 2386-2396. [Google Scholar] [Crossref]
8. Yang, Yi Isaac, et al. "Enhanced sampling in molecular dynamics." *J Chem Phys.* 151.7 (2019). [Google Scholar] [Crossref]
9. Bussi, Giovanni & Alessandro Laio. "Using metadynamics to explore complex free-energy landscapes." *Nat Rev Phys.* 2.4 (2020): 200-212. [Google Scholar] [Crossref]
10. Plimpton,et al. "Computational aspects of many-body potentials." *MRS Bull.* 37.5 (2012): 513-521. [Google Scholar] [Crossref]
11. Abraham, Mark James, et al. "GROMACS: High performance molecular simulations through multi-level parallelism from laptops to supercomputers." *SoftwareX* 1 (2015): 19-25. [GoogleScholar] [CrossRef]
12. Phillips, James C., et al. "Scalable molecular dynamics on CPU and GPU architectures with NAMD." *J. chem. phys.* 153.4 (2020).[GoogleScholar][CrossRef]
13. Cygan, Randall T., et al. "Molecular models and simulations of layered materials." *J Mater Chem* 19.17 (2009): 2470-2481.[Google Scholar] [Crossref]
14. J.A. Anderson, J.G., S.C. Glotzer, Comput. HOOMD website. Mater. Sci. 173 (2002) 109363 2021; Available from: http://glotzerlab.engin.umich.edu/hoomd-blue
15. DCS Computing website. 2021; Available from: https://www.aspherix-dem.com.
16. Trott, Christian, et al. "The kokkos ecosystem: Comprehensive performance portability for high performance computing." *Comput Sci Eng.* 23.5 (2021): 10-18. [GoogleScholar] [CrossRef]
17. Brooks, Bernard R., et al. "CHARMM: the biomolecular simulation program." *J. comput. chem.* 30.10 (2009): 1545-1614. [GoogleScholar] [CrossRef]
18. Páll, Szilárd, et al. "Tackling exascale software challenges in molecular dynamics simulations with GROMACS." nth ed. Springer International Publishing, 2015. EASC 2014, Stockholm, Sweden, April 2-3, 2014,[GoogleScholar] [CrossRef]
19. Smith, W., et al. "The DL_POLY molecular simulation package." CCLRC. Daresbury Laboratory, Daresbury, Warrington, England (1999). [GoogleScholar] [CrossRef]
20. Thompson, Aidan P., et al. "LAMMPS-a flexible simulation tool for particle-based materials modeling at the atomic, meso, and continuum scales." Comput Phys Commun. 271 (2022): 108171.[GoogleScholar] [CrossRef]
21. D'Amico, Marco, & Julita Corbalan Gonzalez. "Energy hardware and workload aware job scheduling towards interconnected HPC environments." *EEE Trans. Parallel Distrib Syst* (2021). [GoogleScholar] [CrossRef]
22. Meloni, Simone, et al. "Efficient particle labeling in atomistic simulations." *J. chem. phys.* 28;126.(2007).[GoogleScholar] [CrossRef]
23. Tuckerman, M. et al. "Reversible multiple time scale molecular dynamics." *J. chem. phys.* 97.3 (1992): 1990-2001. [GoogleScholar] [CrossRef]
24. in't Veld, Pieter J. et al. "Accurate and efficient methods for modeling colloidal mixtures in an explicit solvent using molecular dynamics." *Comput Phys Commun.* 179.5 (2008): 320-329.
25. Shire, Tom, et al. "DEM simulations of polydisperse media: efficient contact detection applied to investigate the quasi-static limit." *Comput Part Mech.* 8.4 (2021): 653-663. [GoogleScholar] [CrossRef]
26. Hockney, Roger W.,et al. *Computer simulation using particles*. crc Press, 2021.[GoogleScholar] [CrossRef]
27. York, Darrin M., et al. "The effect of long-range electrostatic interactions in simulations of macromolecular crystals: A comparison of the Ewald and truncated list methods." *J. chem. phys.* 99.10 (1993): 8345-8348. [GoogleScholar] [CrossRef]
28. Deserno, Markus, & Christian Holm. "How to mesh up Ewald sums. I. A theoretical and numerical comparison of various particle mesh routines." *J. chem. phys.* 109.18 (1998): 7678-7693. [GoogleScholar] [CrossRef]
29. Sutmann, Godehard. "ScaFaCoS–A Scalable library of Fast Coulomb Solvers for particle Systems." (2014).[GoogleScholar]
30. Milestone, William, et al. "Monte Carlo transport analysis to assess intensity dependent response of a carbon-doped GaN photoconductor." *Journal of Applied Physics* 129.19 (2021).[GoogleScholar] [CrossRef]
31. Plimpton, Steve, et al. "Particle-Mesh Ewald and rRESPA for Parallel Molecular Dynamics Simulations." *PPSC*. 1997. [GoogleScholar] [Crossref]
32. Cerdà, Juan J., et al. "P3M algorithm for dipolar interactions." *The J. chem. phys.* 129.23 (2008). [GoogleScholar] [CrossRef]
33. Isele-Holder., et al. "Development and application of a particle-particle particle-mesh Ewald method for dispersion interactions *J. chem. phys.* 137.17 (2012). [GoogleScholar] [CrossRef]
34. Moore, Stan G., & Paul S. Crozier. "Extension and evaluation of the multilevel summation method for fast long-range electrostatics calculations." *J. chem. phys.* 140.23 (2014). [GoogleScholar] [Crossref]
35. Duffy, D. M., & A. M. Rutherford. "Including the effects of electronic stopping and electron–ion interactions in radiation damage simulations." *J. chem. phys.:Condens Matter* 19.1 (2006): 016207.[GoogleScholar] [CrossRef]
36. Rutherford, A. M., & D. M. Duffy. "The effect of electron–ion interactions on radiation damage simulations *J. chem. phys.: Condens Matter* 19.49 (2007): 496201. [GoogleScholar] [CrossRef]
37. Phillips, Carolyn et al. "A two-temperature model of radiation damage in α-quartz." *J. chem. phys.* 133.14 (2010). [GoogleScholar] [CrossRef]
38. Maitland, Geoffrey C. "Intermolecular forces: their origin and determination." *(No Title)* (1981). [GoogleScholar] [CrossRef]
39. Gray, C. G. "KE Gubbins Theory of Molecular Fluids, vol. 1." (1984): 587.[GoogleScholar]

40. Sprik, Michiel. "Effective pair potentials and beyond." *Comput. simul. chem. phys.* (1993): 211-259. [GoogleScholar] [CrossRef]
41. Stone, Anthony. *The theory of intermolecular forces*. oUP oxford, 2013. [GoogleScholar] [CrossRef]
42. Allen, Michael P., et al. "Hard convex body fluids." *Adv chem phys.* 86 (1993): 1-166. [GoogleScholar] [CrossRef]
43. Weeks, John D., et al. "Role of repulsive forces in determining the equilibrium structure of simple liquids." *J chem phys.* 54.12 (1971): 5237-5247. [GoogleScholar] [CrossRef]
44. Holm, Christian. "Efficient methods for long range interactions in periodic geometries plus one application." *Comput Soft Matter: Synth Polym Proteins.* 23 (2004): 195-236. [GoogleScholar] [CrossRef]
45. Price, Sarah L. "Toward more accurate model intermolecular potentials for organic molecules." *Rev Comput Chem.* 14 (2000): 225-289. . [GoogleScholar] [CrossRef]
46. Allinger, Norman L., et al. "Molecular mechanics. The MM3 force field for hydrocarbons. 1." *J Am Chem Soc.* 111.23 (1989): 8551-8566. [GoogleScholar] [CrossRef]
47. Lii, Jenn Huei, & Norman L. Allinger. "Molecular mechanics. The MM3 force field for hydrocarbons. 2. Vibrational frequencies and thermodynamics." *J Am Chem Soc.* 111.23 (1989): 8566-8575. [GoogleScholar] [CrossRef]
48. Lii, Jenn Huei, et al. "Molecular mechanics. The MM3 force field for hydrocarbons. 3. The van der Waals' potentials and crystal data for aliphatic and aromatic hydrocarbons." *J Am Chem Soc.* 111.23 (1989): 8576-8582. [GoogleScholar] [CrossRef]
49. Allinger, Norman L., et al. "An improved force field (MM4) for saturated hydrocarbons." *J comput chem.* 17.5-6 (1996): 642-668. [GoogleScholar] [CrossRef]
50. Nevins, Neysa, et al. "Molecular mechanics (MM4) calculations on alkenes." *J comput chem.* 17.5-6 (1996): 669-694. [GoogleScholar] [CrossRef]
51. Nevins, Neysa, et al. "Molecular mechanics (MM4) calculations on conjugated hydrocarbons." *J. Comput. Chem.* 17.5-6 (1996): 695-729. [Google Scholar] [Crossref]
52. Weiner, Scott J., et al. "A new force field for molecular mechanical simulation of nucleic acids and proteins." *J. Am. Chem. Soc.* 106.3 (1984): 765-784. [Google Scholar] [Crossref]
53. Cornell, Wendy D., et al. "A second generation force field for the simulation of proteins, nucleic acids, and organic molecules." *J. Am. Chem. Soc.* 117.19 (1995): 5179-5197. [Google Scholar] [Crossref]
54. Brooks, Bernard R., et al. "CHARMM: a program for macromolecular energy, minimization, and dynamics calculations." *J. comput. chem.* 4.2 (1983): 187-217. [Google Scholar] [Crossref]
55. Jorgensen, William L et al. "Development and testing of the OPLS all-atom force field on conformational energetics and properties of organic liquids." *J. Am. Chem. Soc.* 118.45 (1996): 11225-11236. [Google Scholar] [Crossref]
56. Allen, Michael P. & Dominic J. Tildesley. "Computer simulation of liquids." *Clarendon-0.12* (1987). [Google Scholar]
57. Hairer, Ernst, et al."Geometric numerical integration illustrated by the Störmer–Verlet method." *Acta numer.* 12 (2003): 399-450. [Google Scholar] [Crossref]
58. Cotter, Colin John, & Sebastian Reich. "Time stepping algorithms for classical molecular dynamics." *Comput. Nanotechnol., Am. Sci. Publ., appear* (2004). [Google Scholar]
59. Leimkuhler, Benedict, & Sebastian Reich. Simulating hamiltonian dynamics. No. 14. *Camb. univ. press*, 2004. [Google Scholar]
60. Verlet, Loup. "Computer" experiments" on classical fluids. II. Equilibrium correlation functions." *Phys. Rev.* 165.1 (1968): 201. [Google Scholar] [Crossref]
61. Moritz, Helmut. "Introduction to classical mechanics." *Theory of Satellite Geodesy and Gravity Field Determination*. Berl. Heidelb.: Springer Berl. Heidelb., 2005. 9-68.. [Google Scholar] [Crossref]
62. de Leeuw, Simon W et al. "Hamilton's equations for constrained dynamical systems." *J. stat. phys.* 61 (1990): 1203-1222. [Google Scholar] [Crossref]
63. Andersen, Hans C. "Rattle: A "velocity" version of the shake algorithm for molecular dynamics calculations." *J. comput. Phys.* 52.1 (1983): 24-34. [Google Scholar] [Crossref]
64. Gay, J. G., & B. J. Berne. "Modification of the overlap potential to mimic a linear site–site potential." *J. Chem. Phys.* 74.6 (1981): 3316-3319. [Google Scholar] [Crossref]
65. Price, S. L. et al. "Explicit formulae for the electrostatic energy, forces and torques between a pair of molecules of arbitrary symmetry." *Mol. Phys.* 52.4 (1984): 987-1001.[Google Scholar] [Crossref]
66. Arnol'd, Vladimir Igorevich. *Mathematical methods of classical mechanics*. Vol. 60. *Springer Sci. Bus. Media*, 2013.[Google Scholar]
67. Bozorgpour, Reza, et al. "Exploring the Role of Molecular Dynamics Simulations in Most Recent Cancer Research: Insights into Treatment Strategies." arXiv preprint arXiv:2310.19950 (2023).[Google Scholar] [Crossref]
68. Fu, S-P., et al. "Lennard-Jones type pair-potential method for coarse-grained lipid bilayer membrane simulations in LAMMPS." *Comput. Phys. Commun.* 210 (2017): 193-203.[Google Scholar] [Crossref]
69. Tan, Jifu, et al . "A parallel fluid–solid coupling model using LAMMPS and Palabos based on the immersed boundary method." *J. comput. sci.* 25 (2018): 89-100.[Google Scholar] [Crossref]
70. Savan, Vachhani, et al. "Multiscale Analysis of Bulk Metallic Glasses for Cardiovascular Applications." *Adv. Eng. Des.: Sel. Proc. FLAME 2018, Springer Singap*. [Google Scholar] [Crossref]
71. Ye, Huilin, Zhiqiang Shen, & Ying Li. "Computational modeling of magnetic particle margination within blood flow through LAMMPS." *Comput. Mech.* 62 (2018): 457-476.[Google scholar] [Crossref]
72. Oyewande, O. E., O. D. Olabiyi, & M. L. Akinyemi. "Molecular dynamics simulations and ion beam treatment of polyethylene." *J. Phys. Conf. Ser.* Vol. 1299. No. 1. IOP Publishing, 2019.[Google scholar] [Crossref]
73. Ye, Huilin, Ying Li, & Teng Zhang. "Magttice: A lattice model for hard-magnetic soft materials." *Soft Matter* 17.13 (2021): 3560-3568. [Google scholar] [Crossref]
74. Tranchida, Julien. "LAMMPS, a brief overview and examples of modifications." (2021).[Google scholar]
75. Ghamari, Nakisa, et al. "Evaluation of the Interaction of Curcumin and Nigella Sativa on Brain Antitumor Molecule Using an Equilibrium Dynamics Simulation Tool for Biomedical Applications." *Nano Biomed. Eng.* 14.4 (2022).[Google scholar][Crossref]
76. Gowthaman, S. "A review on mechanical and material characterisation through molecular dynamics using large-scale atomic/molecular massively parallel simulator (LAMMPS)." *Funct. Compos. Struct.* 5.1 (2023): 012005.[Google scholar] [Crossref]
77. Chávez Thielemann, Hernán, et al. "From GROMACS to LAMMPS: GRO2LAM: A converter for molecular dynamics software." *J. mol. model*. 25 (2019): 1-12.[Google scholar] [Crossref]
78. Falick & Ahyo Chang. Vertex Modeling of Confluent Cellular Networks Within a Molecular Dynamics Framework in LAMMPS. Diss. University of Colorado at Boulder, 2023.[Google scholar]
79. Panah, et al. "Molecular Dynamics Simulation of Atomic Interaction between Main Protein of Human Prostate Cancer and Fe/C720 Buckyballs-Statin Structures." (2022).[Google scholar] [Crossref]
80. Evans, T. J., et al. "Porting the AMBER forcefield to LAMMPS– massively parallel molecular dynamics simulations of DNA." [Google scholar]
81. 81.AlDosari, Sahar Mohammed, et al. "Drug release using nanoparticles in the cancer cells on 2-D materials in order to target drug delivery: A numerical simulation via molecular dynamics method." *Eng. Anal. Bound. Elem*. 148 (2023): 34-40.[Google scholar] [Crossref]
82. Tsukanov, Aleksey Alekseevich, & Sergey Grigorievich Psakhie. "Two-dimensional Al hydroxide interaction with cancerous cell membrane building units: Complexed free energy and orientation analysis." *AIP Conf. Proc*.,Vol. 1882. No. 1. AIP Publishing, 2017.[Google scholar] [Crossref]
83. Jeevanandam, Jaison, et al. "Advancing aptamers as molecular probes for cancer theranostic applications—the role of molecular dynamics simulation." *Biotechnol. j.* 15.3 (2020): 1900368.[Google scholar] [crossref]
84. Kuninty, Praneeth R., et al. "Cancer immune therapy using engineered' tail-flipping'nanoliposomes targeting alternatively activated macrophages." *Nat. commun.* 13.1 (2022): 4548.[Google scholar] [Crossref]
85. Faizi, Atousa, et al. "Drug delivery by SiC nanotubes as nanocarriers for anti-cancer drugs: investigation of drug encapsulation and system stability using molecular dynamics simulation." *Mater. Res. Express* 8.10 (2021): 105012.[Google scholar] [Crossref]
86. Johnstone, Sarah E., et al. "Large-scale topological changes restrain malignant progression in colorectal cancer." *Cell* 182.6 (2020): 1474-1489.[Google scholar] [Crossref]
87. Misra, Santosh K., et al. "Pro-nifuroxazide self-assembly leads to triggerable nanomedicine for anti-cancer therapy." *ACS appl. mater. interfaces* 11.20 (2019): 18074-18089.[Google scholar] [Crossref]

88. Tohidi, Shabnam, and Mehrdad Aghaie-Khafri. "Chitosan-Coated MIL-100 (Fe) as an Anticancer Drug Carrier: Theoretical and Experimental Investigation." *ACS Med. Chem. Lett.* 14.9 (2023): 1242-1249.[Google scholar] [Crossref]
89. Zhang, Xiaoyu, et al. "Large Laser Spot-Swift Mapping Surface-Enhanced Raman Scattering on Ag Nanoparticle Substrates for Liquid Analysis in Serum-Based Cancer Diagnosis." *ACS Appl. Nano Mater.* 5.10 (2022): 15738-15747.[Google scholar] [Crossref]
90. Xu, Jianchang, et al. "Copolymer-functionalized cellulose nanocrystals as a pH-and NIR-triggered drug carrier for simultaneous photothermal therapy and chemotherapy of cancer cells." *Biomacromolecules* 23.10 (2022): 4308-4317.[Google scholar] [Crossref]
91. Xu, Li, et al. "Supramolecular cyclic dinucleotide nanoparticles for STING-Mediated Cancer Immunotherapy." *ACS nano* 17.11 (2023): 10090-10103.[Google Scholar] [Crossref]
92. Rossinelli, Diego, et al. "The in-silico lab-on-a-chip: petascale and high-throughput simulations of microfluidics at cell resolution." *Proc. Int. Conf. High Perform. Comput. Netw. Storage Anal.,* 2015.[Google Scholar] [Crossref]
93. Mercan, Kadir, and Ömer Civalek. "Comparative Stability Analysis of Boron Nitride Nanotube using MD Simulation and Nonlocal Elasticity Theory." *Int. J. Eng. Appl. Sci.* 13.4 (2022): 189-200.[Google Scholar] [Crossref]
94. Yarahmadi, Mehran, & Sadollah Ebrahimi. "The effect of type and concentration of functional groups on the molecular adsorption of paclitaxel onto graphene oxide in the aqueous environments using molecular dynamics simulations." *Sci. J. Kurd. Univ. Med. Sci.* 24.4 (2019): 138-148.[Google Scholar]
95. Bahreini, Maziar, and Arezoo Ghaffari. "Computational Study of Diol Camptothecin Drug Delivery Process Using MPEG-1 based Nanosome Structure: Molecular Dynamics Approach." (2023) [Google Scholar] [Crossref]
96. Roosta, Sara, et al. "Molecular dynamics simulation study of boron-nitride nanotubes as a drug carrier: from encapsulation to releasing." *RSC adv.* 6.11 (2016): 9344-9351.[Google Scholar] [Crossref]
97. DeLong, Robert K., et al. "Comparative molecular immunological activity of physiological metal oxide nanoparticle and its anticancer peptide and rna complexes." *Nanomaterials* 9.12 (2019): 1670.[Google scholar] [Crossref]
98. Zarghami Dehaghani, Maryam, et al. "Theoretical encapsulation of fluorouracil (5-FU) anti-cancer chemotherapy drug into carbon nanotubes (CNT) and boron nitride nanotubes (BNNT)." *Molecules* 26.16 (2021): 4920.[Google Scholar] [Crossref]
99. Yang, Hung-Wei, et al. "Aptasensor designed via the stochastic tunneling-basin hopping method for biosensing of vascular endothelial growth factor." *Biosens. Bioelectron.* 119 (2018): 25-33.[Google Scholar] [Crossref]
100. Hatam-Lee, S. Milad, et al. "Interfacial thermal conductance between gold and SiO2: A molecular dynamics study." *Nanoscale Microscale Thermophys. Eng.* 26.1 (2022): 40-51.[Google scholar] [Crossref]
101. Verlackt, C. C. W., et al. "Atomic-scale insight into the interactions between hydroxyl radicals and DNA in solution using the ReaxFF reactive force field." *New J. Phys.* 17.10 (2015): 103005.[Google Scholar] [Crossref]
102. Shahbazi, Fatemeh, et al. "A Molecular Dynamics Model for Biomedical Sensor Evaluation: Nanoscale Numerical Simulation of an Aluminum-Based Biosensor." *2022 44th Annu. Int. Conf. IEEE Eng. Med. Biol. Soc. (EMBC)*. IEEE, 2022.[Google Scholar] [Crossref]
103. Tsukanov, Alexey A. & Sergey G. Psakhie. "Adsorption of charged protein residues on an inorganic nanosheet: Computer simulation of LDH interaction with ion channel." *AIP Conf. Proc.* Vol. 1760. No. 1. AIP Publishing, 2016.[Google Scholar] [Crossref]
104. Tsukanov, Alexey A., & Sergey G. Psakhie. "A molecular dynamic study of charged nanofilm interaction with negative lipid bilayer." *AIP Conf. Proc., Vol, 1623, No, 1. Am. Inst. Phys.*, 2014.[Google Scholar] [Crossref]
105. MacKerell Jr, Alex D., et al"All-atom empirical potential for molecular modeling and dynamics studies of proteins." *j. phys. chem. B* 102.18 (1998): 3586-3616.[Google Scholar] [Crossref]
106. Zhang, Lili, et al. "Molecular dynamics simulations of heterogeneous cell membranes in response to uniaxial membrane stretches at high loading rates." *Sci. rep.* 7.1 (2017): 8316.[Google scholar] [Crossref]
107. Lequieu, Joshua, et al. "1CPN: A coarse-grained multi-scale model of chromatin." *J. Chem. Phys.* 150.21 (2019).[Google Scholar] [Crossref]
108. Siani, Paulo, and Cristiana Di Valentin. "Effect of dopamine-functionalization, charge and pH on protein corona formation around TiO 2 nanoparticles." *Nanoscale* 14.13 (2022): 5121-5137.[Google Scholar] [Crossref]
109. da Rocha, Edroaldo Lummertz, Giovanni Finoto Caramori, and Carlos Renato Rambo. "Nanoparticle translocation through a lipid bilayer tuned by surface chemistry." *Phys. chem. chem. phys.* 15.7 (2013): 2282-2290.[Google Scholar] [Crossref]
110. Bao, Weier, et al. "Experimental and theoretical explorations of nanocarriers' multistep delivery performance for rational design and anticancer prediction." *Sci. Adv.* 7.6 (2021): eaba2458.[Google Scholar] [Crossref]
111. Mohebali, M., et al. "An MD-based systematic study on the mechanical characteristics of a novel hybrid CNT/graphene drug carrier." *J. Mol. Model.* 26 (2020): 1-12.[Google Scholar] [Crossref]
112. Zhang, Liang, et al. "Molecular dynamics simulations of the microscopic mechanism of thermal conductivity enhancement of ethylene Glycol–Water–Au nanofluids." *Appl. Surf. Sci.* 609 (2023): 155389.[Google scholar] [Crossref]
113. Abadi, Rouzbeh, et al. "Computational modeling of graphene nanopore for using in DNA sequencing devices." Phys. E: Low-dimens. Syst. Nanostructures 103 (2018): 403-416.[Google scholar] [Crossref]
114. Siani, Paulo, et al. "Molecular dynamics simulations of cRGD-conjugated PEGylated TiO2 nanoparticles for targeted photodynamic therapy." J. Colloid Interface Sci. 627 (2022): 126-141.[Google scholar] [Crossref]
115. Rebocho, Tatiana Sofia Carvalho. Modeling cell adhesions and proliferation in complex environments. Diss. 2023.[Google scholar]
116. Cunha, Ana F., et al. "Cell response in free-packed granular systems." *ACS Appl. Mater. Interfaces* 14.36 (2022): 40469-40480.[Google scholar] [Crossref]
117. sereshki, s. and S. Lonardi, Predicting Differentially Methylated Cytosines in TET and DNMT3 Knockout Mutants via a Large Language Model. bioRxiv, 2024: p. 2024.05. 02.592257.
118. Sereshki, S., et al., On the prediction of non-CG DNA methylation using machine learning. NAR genomics and bioinformatics, 2023. 5(2): p. lqad045.
119. Maher, S.P., et al., A drug repurposing approach reveals targetable epigenetic pathways in Plasmodium vivax hypnozoites. bioRxiv, 2023